\documentclass[]{aa}
\usepackage{graphicx,amsmath}
\usepackage{txfonts}
\usepackage{color,ulem}
\newcommand{\figref}[1]{Fig.\,\ref{#1}}
\newcommand{\brafracket}[2]{\left( \frac{#1}{#2} \right)}

\newcommand{\sub }[1]{_{\mathrm{#1}}}
\def\wig#1{\mathrel{\hbox{\hbox to 0pt{%
          \lower.6ex\hbox{$\sim$}\hss}\raise.4ex\hbox{$#1$}}}}
 \newcommand{\dd}{\mathrm{d}}
\newcommand{\Mstar}{M_{\star}}
\newcommand{\Mfin}{M\sub{final}}
\newcommand{\mke}{m\sub{ke}}
\newcommand{\MCZ}{M\sub{CZ}}
\newcommand{\MCZeff}{M\sub{CZ}^{\mathrm{eff}} }
\newcommand{\mzp}{M\sub{Z,p}}
\newcommand{\dMO}{\Delta M_0}
\newcommand{\Zs}{Z\sub{surf}}
\newcommand{\Za}{Z\sub{acc}}
\newcommand{\Mrocktot}{M\sub{rock,tot}}
\newcommand{\Mrockreq}{M\sub{rock,req}}
\newcommand{\Zrockacc}{Z\sub{rock,acc}}
\newcommand{\Zrockini}{Z\sub{rock,ini}}

\newcommand{\mdr}{M\sub{disk,rock}}
\newcommand{\Zini}{\mathrm{Z_{ini}}}
\newcommand{\Zsun}{\mathrm{Z_{\odot}}}
\newcommand{\dM}{\Delta M}
\newcommand{\fir}{f\sub{ice/rock}}
\newcommand{\teff}{T\sub{eff}}
\newcommand{\tp}{t\sub{p}}
\newcommand{\dfeh}{\Delta{\mathrm{[Fe/H]}}}
\newcommand{\Msun}{{\mathrm M_\odot}}
\newcommand{\Mjup}{\mathrm{M_{Jup}}}
\newcommand{\Rsun}{{\mathrm R_\odot}}
\newcommand{\Lsun}{{\mathrm L_\odot}}
\newcommand{\Mearth}{{\mathrm M_\oplus}}
\defcitealias{Kunitomo+17}{Paper I}
\usepackage[colorlinks = true,
        linkcolor = blue,
        urlcolor  = blue,
        citecolor = blue,
        anchorcolor = blue
]{hyperref}
\bibpunct{(}{)}{;}{a}{}{,} 

\begin{document}

   \title{Revisiting the pre-main-sequence evolution of stars}
   \subtitle{II. Consequences of planet formation on stellar surface composition\thanks{
   The evolutionary models are available at the CDS via anonymous ftp to cdsarc.u-strasbg.fr (***.**.***.*) or via http://cdsarc.u-strasbg.fr/viz-bin/qcat?J/A+A/***/***
   }}

        \titlerunning{Consequences of planet formation on stellar surface composition}
        \authorrunning{M. Kunitomo, T. Guillot, S. Ida and T. Takeuchi}

   \author{Masanobu Kunitomo\inst{\ref{inst1}, \ref{inst2}},
          Tristan Guillot\inst{\ref{inst3}},
          Shigeru Ida\inst{\ref{inst4}}
          \and
          Taku Takeuchi\inst{\ref{inst5}}
          \fnmsep\thanks{Present affiliation: Sanoh Industrial Co., Ltd., Japan.} 
          }

   \institute{Department of Earth and Planetary Science, 
                The University of Tokyo, 
                7-3-1, Hongo, Bunkyo-ku, Tokyo 113- 0033, Japan\label{inst1}\\
                \email{kunitomo@eps.s.u-tokyo.ac.jp}
                \and
                Department of Physics, 
                Nagoya University, Furo-cho, Chikusa-ku, Nagoya, 
                Aichi 464-8602, Japan\label{inst2}
                \and
                Laboratoire Lagrange, UMR 7293,
                Universit\'e C\^ote d'Azur,
                CNRS,
                Observatoire de la C\^ote d'Azur,
                06304 Nice CEDEX 04, France\label{inst3}
                \and
                         Earth-Life Science Institute,
                         Tokyo Institute of Technology,
                         2-12-1 Ookayama, Meguro-ku, Tokyo 152-8551, Japan\label{inst4}
         \and
                         Department of Earth and Planetary Sciences, 
                         Tokyo Institute of Technology,
                         2-12-1 Ookayama, Meguro-ku, Tokyo 152-8551, Japan\label{inst5}
}
   \date{Received 29 March 2018; accepted 13 August 2018}

 
  \abstract
     {}
   {
We want to investigate how planet formation is imprinted on stellar surface composition using up-to-date stellar evolution models.
   }
   {
   We simulate the evolution of pre-main-sequence stars as a function of
   the efficiency of heat injection during accretion, the deuterium mass fraction, and the stellar mass, $\Mstar$.
   For simplicity, we assume that planet formation leads to the late accretion of zero-metallicity gas, diluting the surface stellar composition as a function of the mass of the stellar outer convective zone. We estimate that in the solar system, between $97$ and $168\,\Mearth$ of condensates formed planets or were ejected from the system. 
   We adopt $150\,{\Mearth}(\Mstar/\Msun)(Z/\Zsun)$ as an uncertain but plausible estimate of the mass of heavy elements that is not accreted by stars with giant planets, including our Sun. 
        By combining our stellar evolution models to these estimates, we evaluate the consequences of planet formation on stellar surface composition.
   }
   {
     We show that after the first $\sim0.1$ Myr during which stellar structure can differ widely from the usually assumed fully convective structure, the evolution of the convective zone follows classical pre-main-sequence evolutionary tracks within a factor of two in age.
        We find that planet formation should lead to a scatter in stellar surface composition that is larger for high-mass stars than for low-mass stars. 
        We predict a spread in [Fe/H] of approximately {0.05\,dex} for stars with a temperature of $\teff\sim 6500\,$K, to $0.02$\,dex for stars with $\teff\sim 5500\,$K, marginally compatible with differences in metallicities observed in some binary stars with planets.  Stars with $\teff\ge 7000\,$K may show much larger [Fe/H] deficits, by 0.6\,dex or more, in the presence of efficient planet formation, compatible with the existence of  refractory-poor $\lambda$ Boo stars.
        We also find that planet formation may explain the lack of refractory elements seen in the Sun as compared to solar twins, but only if the ice-to-rock ratio in the solar-system planets is less than $\approx0.4$ and planet formation began less than $\approx1.3$\,Myr after the beginning of the formation of the Sun. 
   }
   {}


   \maketitle
%

\section{Introduction} \label{sec:intro}

How does planet formation affect stellar composition?
Most of the gas in protoplanetary disks eventually accretes onto the host star within several million years \citep[e.g.,][]{Haisch+01}. 
Materials that will eventually form planets must condense during this phase.
Recent meteoritic evidence indicates that Jupiter's core should have formed rapidly, within {1\,Myr} \citep{Kruijer+17}.
Disk materials accreted by the star after this time must have been poor in heavy elements\footnote{
We use ``heavy elements'' to describe all species able to condense and separate from hydrogen and helium  in the protoplanetary disk.}
because of the gap created by the growing giant planets \citep[e.g.,][]{Paardekooper+Mellema04, Guillot+14, Morbidelli+16}.
The accretion of such a diluted gas must result in a refractory-poor composition of the stellar surface. 
Hereafter, we refer to this mechanism as \textit{dilution}.
On the other hand, even after disk dispersal, the formation of planets can lead to the ingestion by the central star of planetesimals or even possibly planets, therefore yielding an increase in its metallicity \citep[e.g.,][]{Spina+14a,Tognelli+16}. Hereafter we refer to this mechanism as \textit{pollution}.

\begin{figure}[!t]
  \begin{center}
        \includegraphics[width=7.4cm,keepaspectratio]{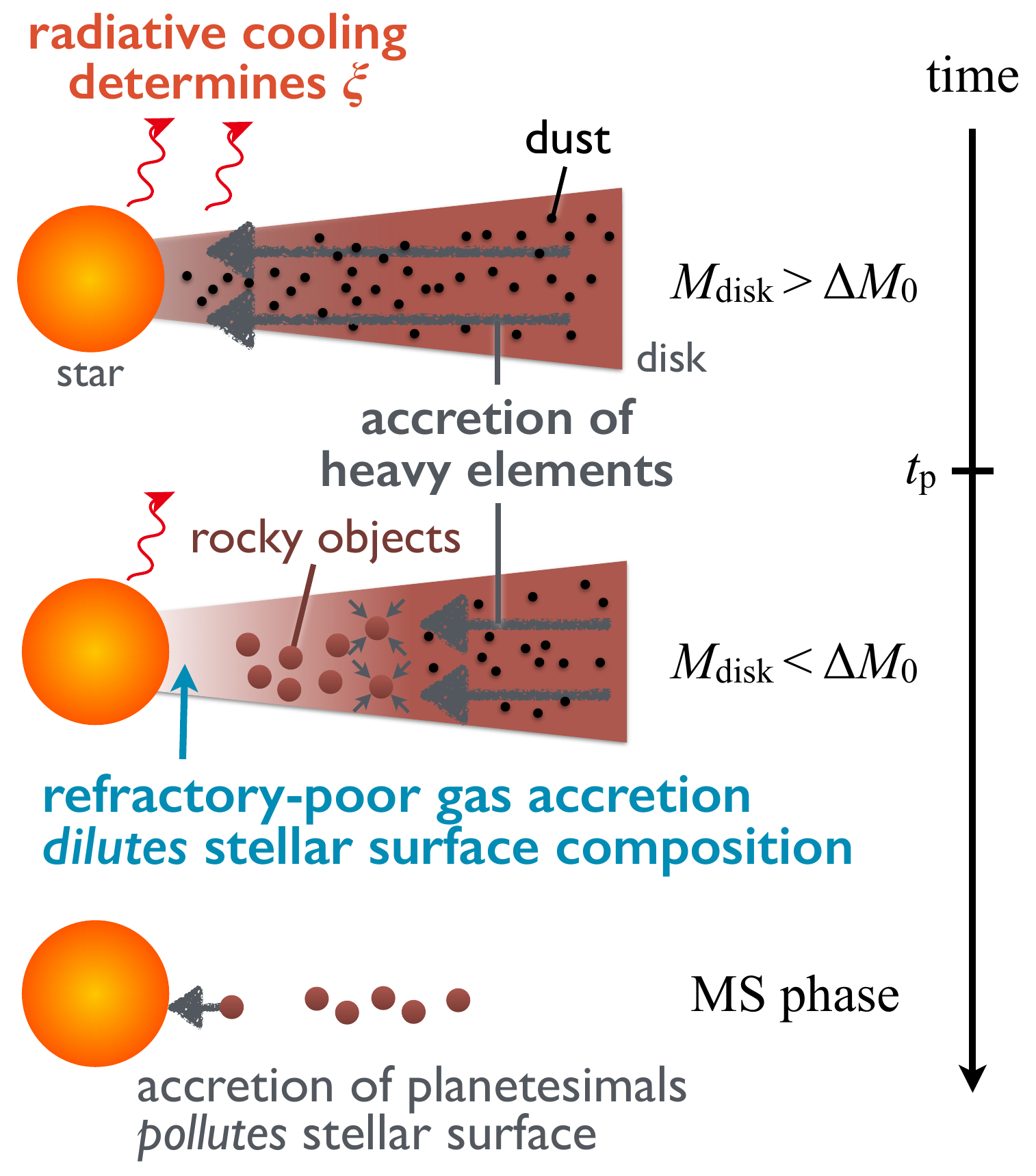}
        \caption{\small{
        Schematic diagram illustrating how planet formation modifies the composition of the stellar surface.
A star first accretes hydrogen, helium and heavy elements (equivalently, metals) from a protoplanetary disk of mass $M\sub{disk}(t)$. 
After a time $\tp$, planet formation removes refractory elements from the accreting gas flow, leading to a dilution of the stellar surface.
At this point, the mass in the gas disk (to be eventually accreted by the star) is $\dMO\equiv M\sub{disk}(\tp)$.
        Later, after disk dispersal, the accretion of rocky objects (e.g., planetesimals) can potentially increase the stellar metallicity (pollution).
        }}\label{fig:pollution}
    \end{center}
\end{figure}

In solar-type stars, an outer convective zone is present and rapidly mixes the stellar {surface}. The process is extremely fast ($\sim$years) compared to evolution timescales and can be considered as instantaneous. The magnitude of the dilution depends both on the amount of heavy elements retained to form planets and on the extent of the outer convective zone of the star.

Figure\,\ref{fig:pollution} illustrates how planet formation may affect stellar surface composition and shows the model that we consider in this article.
The star first accretes gas with a composition that is equal to that of the molecular cloud core (mass fraction of heavy elements $\Zini$). When planetesimals begin to form, refractory-poor ($<\Zini$) gas is accreted. 
Accretion of planetesimals and planets then proceeds mostly later when the star is fully formed and the gas disk has dispersed. 
The crucial quantities involved are 
(i) $M\sub{disk}(t)$, the mass of the disk as a function of time $t$, 
(ii) $\MCZ(t)$, the mass of the outer convective zone, 
and (iii) $\tp$ the time at which planetesimals begin to form. In addition we define $\dMO$ as the mass of the disk at time $\tp$.

Circumstellar disks are short-lived, a few to 10 Myr at most {\citep[e.g.,][]{Mamajek09}}. This implies that the dilution mechanism takes place while the central star is still on the  pre-main-sequence (pre-MS) phase (for stars $\la1.5\,\Msun$). Contrary to what happens later on the main sequence (MS), on the early pre-MS, even relatively massive stars (say, $1.6\,\Msun$) possess a relatively thick outer convective zone. This convective zone shrinks with age but on a timescale of a few to tens of millions of years. In the classical picture, a $1\,\Msun$ star at 10\,Myr still has $\MCZ\approx0.4\,\Msun$ {\citep[see, e.g.,][]{Iben65,DAntona+Mazzitelli94,Stahler+Palla05}}.
However, pre-MS evolution has recently been revisited \citep{BCG09,BVC12,Baraffe+17,Hosokawa+11,Vorobyov+17} and found to be more complex than previously thought: {in} particular, it appears to be sensitive to the entropy of the gas accreted by the star.
\citet[][]{BC10} and \citet{Tognelli+15} showed that in this case, the evolution of the stellar internal structure may be significantly different from the classical picture (see Sect.\,\ref{sec:int-evols}).

In \citet[][hereafter \citetalias{Kunitomo+17}]{Kunitomo+17}, we found that the pre-MS evolution is also sensitive to the deuterium abundance,  in particular in the low-entropy accretion cases \citep[see also][]{Stahler88, Hosokawa+Omukai09, Tognelli+15}. Moreover we constrained possible accretion scenarios from the comparison with effective temperature--luminosity relations in young clusters, and showed that significant departures from the classical evolution scenario (a heat accretion efficiency of 10\% or less, see {Sect.\,\ref{sec:constraints}}) should be rare. 

In this article, we investigate the internal structure evolution of pre-MS stars under a variety of settings and evaluate the consequences of planet formation on stellar surface composition.
We compare our predictions with some observations potentially linked to the dilution {and pollution mechanisms}: 
the trends in stellar metallicity versus effective temperatures in the Hyades cluster \citep{Takeda+13, Takeda+17}, the metal-poor surface compositions of $\lambda$ Boo stars {\citep{Kama+15, Murphy+Paunzen17}},
the chemical inhomogeneity in some binary systems,
and the lack of refractory materials measured in our Sun when compared to solar twins {\citep[e.g.,][]{Melendez+09, Ramirez+09, Chambers10}.}

This paper is organized as follows. 
In Sect.\,\ref{sec:background}, we describe the astrophysical background including the evolution of the metallicity of accreting materials.
In Sect.\,\ref{sec:method}, we provide a brief summary of our physical model and of the computation methods. 
Using our up-to-date stellar evolution models, we explore the consequences of planet formation on stellar surface composition in Sect.\,\ref{sec:discussion}.
In Sect.\,\ref{sec:discussion-solar}, we discuss the limits within which the dilution mechanism may explain the anomaly in the solar surface composition compared to solar twins.
The results are summarized in Sect.\,\ref{sec:conclusion}.

\section{Planet formation and accretion onto stars} \label{sec:background}

Stars are formed by the gravitational collapse of molecular cloud cores. 
Due to angular momentum conservation, the contraction of the core results in the formation of both a central stellar seed and a circumstellar disk. The stellar seed accretes mass through the disk in which planets are formed.

Planetesimal and planet formation however is not granted. Grains formed in disks (pebbles) tend to rapidly migrate inward to be accreted by the central star \citep[e.g.,][]{Adachi+76, Weidenschilling77a,Nakagawa+86, Takeuchi+Lin02}. 
We observe nevertheless that most stars host Earth-mass planets, but few of them, of the order of 14\%, host giant planets \citep{Mayor+11}. 
Theoretical models indeed indicate that giant planet formation requires special conditions and is therefore relatively rare \citep{Ida+Lin08V}.
The presence of four giant planets in our solar system appears to be a rare case. 

As soon as a proto-giant planet of $\sim 20\,\Mearth$ appears, it carves a gap in the disk \citep{Paardekooper+Mellema04, Duffell+MacFadyen13, Kanagawa+15}, separating the disk in two and leading to the retention of the migrating grains \citep{Kobayashi+12, Guillot+14, Morbidelli+16, Taki+16}. It is therefore highly probable that the presence of at least four giant planets in our solar system led to the retention of significantly more refractory material into planets than for the majority of stars.  
 
On this basis, we assume that, in Sect.\,\ref{sec:discussion-solar}, $\sim90\%$ of the other solar twins retained, on average, much less refractories into planets than our Sun \citep[][see also Sect.\,\ref{sec:background-solar}]{Ramirez+09}.

We hereafter first evaluate the mass of condensates that must be retained in planetary objects in the early solar system.

\subsection{Total condensates in the early solar system} \label{app:Mrock}
The condensates consist of currently existing objects $(\simeq59$--$102\,\Mearth)$ and ejected or otherwise missing objects $(\simeq38$--${66}\,\Mearth)$.

First, {we can account for the following mass of condensates (heavy elements) in our solar system except the Sun}:
\begin{itemize}
\item $2.01\,\Mearth$ rocks in the four terrestrial planets \citep{Lang92}, 
\item $5\times10^{-4}\,\Mearth$ condensates in the asteroid belt \citep{Lang92}, 
\item 10--40$\,\Mearth$ in Jupiter \citep{Guillot05, Miguel+16, Wahl+17}, 
\item 20--30$\,\Mearth$ in Saturn \citep{Guillot05, Helled+Guillot13}, 
\item 12--13$\,\Mearth$ in Uranus and 13.5--15$\,\Mearth$ in Neptune \citep[estimated from][]{Nettelmann+13},
\item 0.2--0.3$\,\Mearth$ in the Edgeworth-Kuiper belt and scattered disk \citep{Gomes+08}, and
\item 1--1.5$\,\Mearth$ in the Oort cloud \citep{Kaib+Quinn08, Brasser+Morbidelli13}. This number is essentially based on comets coming from the outer Oort cloud, where orbits may be perturbed by neighboring stars. The inner Oort cloud may hide some larger objects, like Planet 9 \citep{Batygin+Brown16}. We account for these below, when including calculations of planetesimals and planets ejected  (or, in this case, nearly-ejected) from the solar system.    
\end{itemize}
The total condensate mass is therefore between 58.7 and 101.8\,$\Mearth$. 

Next, we evaluate the mass of missing objects from the asteroid belt, the Kuiper belt, and the region around Uranus and Neptune.
\citet{OBrien+07} estimate that the primordial asteroid belt has an initial mass of $\sim 2\,\Mearth$. Of these however, about one forth are embryos that end up accreted by planets. This means that about $1.5\,\Mearth$ of asteroid belt material was ejected from the solar system.
The structure of the Kuiper belt and the eccentricities of the giant planets lead (within the Nice model) to a requirement of about $35\,\Mearth$ present in the outer solar system - after the giant planets had formed \citep{Tsiganis+05}. Of this $35\,\Mearth$ , only about $0.3\,\Mearth$ is accreted by the fully formed giant planets \citep{Matter+09}. We therefore assume that $35\,\Mearth$ of planetesimals in the Jupiter-Saturn region was ejected. 
Subsequently, when Uranus and Neptune formed, the region must have contained a considerable mass of planetesimals and protoplanets. \citet{Izidoro+15} ran simulations with initially 30 and $60\,\Mearth$ in embryos, in addition to the fully-formed Jupiter and Saturn.  The results of his simulations \citep[][Izidoro, private communication 2017]{Izidoro+15} are summarized in Table\,\ref{tab:izidoro}. They indicate that between $1.7$ and $29.5\,\Mearth$ of planetesimals was ejected or implanted in the inner Oort cloud, and that the Sun was hit by 0 to $1\,\Mearth$ of planetesimals. 
{In total, we therefore estimate the total mass of condensates (rocks and ices) ejected from the solar system to lie between 38 and $66\,\Mearth$.}

   \begin{table}[!tb]
      \caption[]{Planetesimal mass in $\Mearth$ lost when Uranus and Neptune formed in the simulations by \citet{Izidoro+15}.}
         \label{tab:izidoro} \small
         \begin{tabular}{p{2.1cm}|p{1.6cm}p{0.8cm}p{1.4cm}p{1.1cm}}
            \hline
            \hline
            \noalign{\smallskip}
            Initial total mass in embryos & mean ejected mass & hit the Sun & hit the giant planets & remaining mass\\
            \noalign{\smallskip}
            \hline
            \noalign{\smallskip}
            60  & $\sim13$  & $\sim1$ & $\sim4$--5 & $\sim41$--42  \\
            30  & $\sim0.7$ & $\sim0$ & $\sim0.5$  & $\sim29$  \\
            \noalign{\smallskip}
            \hline
            \noalign{\smallskip}
         \end{tabular} 
          \tablefoot{The remaining mass must be used to form Uranus and Neptune (25.5--$28\,\Mearth$ in heavy elements) and the inner Oort cloud objects, like perhaps planet 9.}
        \normalsize
            \end{table}

As a result, about 97 to $168\,\Mearth$ was extracted from the protosolar disk to form planets and about $1\,\Mearth$ may have been implanted into the Sun during the MS phase.
Hereafter we refer to the total mass of condensates accreted into planets or ejected as $\mzp$.
We adopt $\mzp=150\,\Mearth$ in the early solar system.

The total mass of material accreted by the star after planet formation has begun ($t>\tp$, see Fig.\,\ref{fig:pollution}), $\dMO$, can therefore be estimated to be at least $\mzp/\Zini$, where $\Zini$ is the primordial metallicity of the molecular cloud core. In the simple scenario in which the star accretes $Z=0$ gas after $\tp$ (see Sect.\,\ref{sec:Za} hereafter), $\dMO= \mzp/\Zini$. 
As we see below (e.g., Fig.\,\ref{fig:comp-Zs}), changes in metallicity by the dilution effect are only of the order of 5\%, meaning that we can safely assume that $\Zini\approx \Zsun$.
Given that $\Zsun = 0.015$ in the solar system \citep{Asplund+09}, $\Delta M_{0,\odot}=150\,\Mearth/0.015\simeq0.03\,\Msun$.

\subsection{Total condensates in extrasolar systems} \label{app:Mrock-extra}

Higher-mass $(>1\,\Msun)$ stars seem to harbor more massive planetary systems \citep[e.g.,][]{Lovis+Mayor07,Johnson+10,Omiya+12,Bonfils+13}. 
We therefore extend the relation to stars with different masses and metallicities by scaling $\mzp$ linearly with them; that is, we assume that $\mzp=150\,{\Mearth}(\Mstar/\Msun)(\Zini/\Zsun)$ corresponding to $\dMO=0.03\,\Mstar$.
We note that since a relatively small fraction of stars harbor giant planets, 
this is our limiting case for stars that succeeded in forming multiple giant planets. 
The other limiting case is when no planets form, in which case we consider that $\dMO=0$.

An estimate of pollution effects in exoplanetary systems may be obtained through the analysis of the compositional anomalies in the $\ga 25\%$ of white dwarfs. 
\citet{Mustill+18} found that the destabilization of planetary systems results in an accretion rate of planetesimals of about $\sim10^{-4}\,\Mearth/\mathrm{Myr}$ given that the belt mass is $\sim1\,\Mearth$.

\subsection{Consequences of accretion onto MS stars} \label{sec:MSacc}

Before using a more realistic approach, let us first examine the consequences of either the accretion of {rocky} planetesimals (i.e., the metallicity of accreting materials $Z=1$) or the accretion of $0.03\,\Mstar$ of $Z=0$ gas onto $0.5$--$1.5\,\Msun$ MS stars.
While the former may be viewed as a possibility, the latter is not realistic, the gas disk lifetime ($\sim 3\,$Myr) being much shorter than the time to reach the MS ($\sim 30\,$Myr). However, this assumption has been used in the literature, and it is instructive to consider its consequences.  
We note that the internal structure of MS stars is not affected by formation history as we show below.

Changes in stellar surface composition $\Zs$ because of accretion are evaluated from the simple relation
\begin{equation} \label{eq:Zs-MS}
\Zs=
\frac
{\MCZ\,{\Zini} + \dM\sub{acc}\,\Za}
{\MCZ+\dM\sub{acc}}\,,
\end{equation}
{where $\Za$ and $\dM\sub{acc}$ are the metallicity and mass of the accreting materials, respectively.}

First, let us evaluate the pollution of the stellar {surface} due to the accretion of planetesimals or planets. 
{In this case, it is important to note that} any positive difference between the metallicity of the CZ and that of the deeper radiative zone will be rapidly erased by fingering (i.e., Rayleigh-Taylor) instabilities \citep{Vauclair04, Garaud11}, on a timescale estimated to be of the order of $10$ to $100\,$Myr \citep{Theado+Vauclair12}. 

We consider two cases with the $100\,$Myr timescale of the thermohaline convection: a recent  ($<100\,$Myr) accretion of $1\,\Mearth$ of planetesimals or planets (similar to the solar case), or the steady accretion of $10^{-4}\,\Mearth/$Myr (see Sect.\,\ref{app:Mrock-extra}) for an interval of $100\,$Myr.
The latter case is equivalent to assuming $\dM\sub{acc}=0.01\,\Mearth$ and $\Za=1$ in Eq.\,\eqref{eq:Zs-MS}. 

\begin{figure}[!t]
  \begin{center}
    \includegraphics[width=\hsize,keepaspectratio]{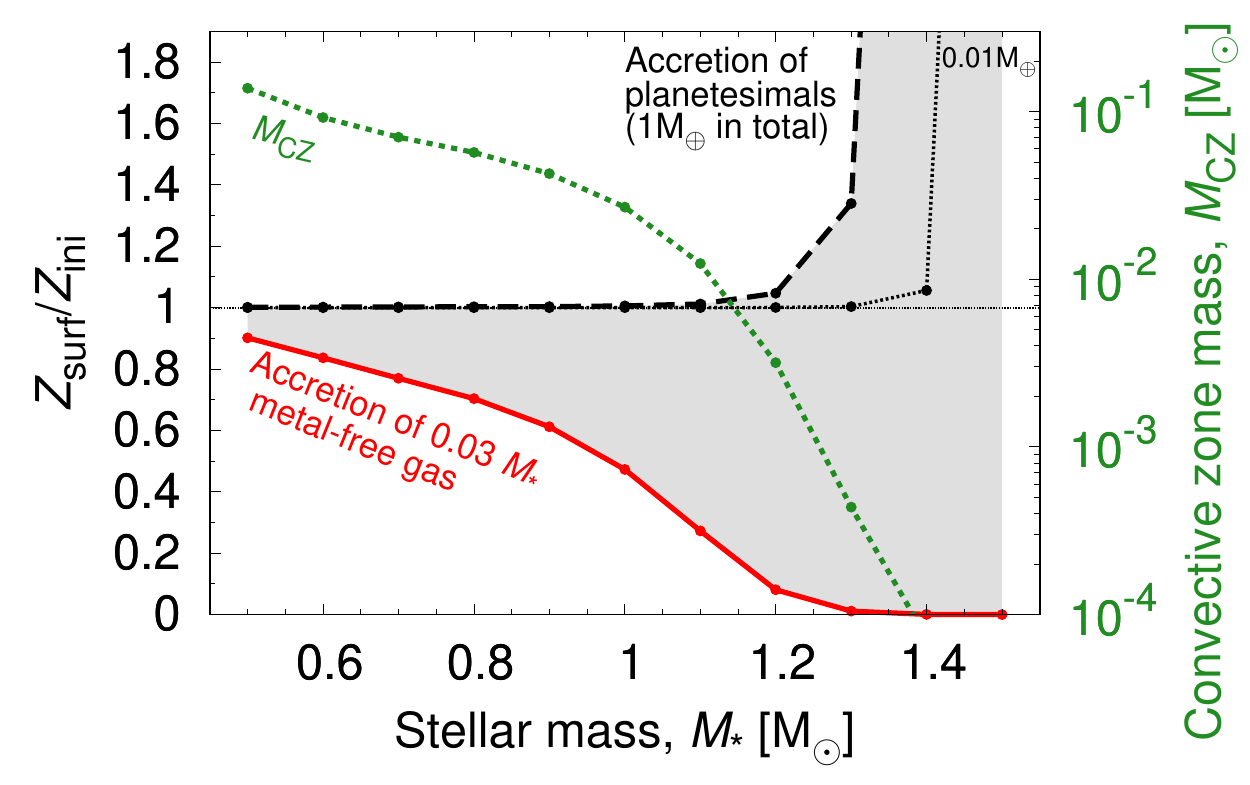}
        \caption{\small
        Potential modification of the surface metallicity of MS stars after the engulfment of $1\,\Mearth$ of planetesimals (dashed line) or the accretion of $0.03\,M_{\star}$ metal-free gas (plain), both resulting as a consequence of the formation of a solar-system analog. 
        {The pollution by $0.01\,\Mearth$ planetesimals is also shown by the thin dotted line.}
        The stellar convective zone mass, $M\sub{CZ}$, is shown as a green dotted line. All values are evaluated at an age of 0.6\,Gyr. The potential changes in composition due to planet formation are indicated in gray.
        }\label{fig:MS-impact}
    \end{center}
\end{figure}

The resulting changes in stellar composition for these two cases are shown in Fig.\,\ref{fig:MS-impact} for MS stars. They are negligibly small for stars with masses lower than $1.2\,\Msun$ but become significant for higher-mass stars because of their thin outer CZ. 
However, even for the high-mass stars, the observed pollution effect in the CZ is likely to be significantly reduced compared to the estimates of Fig.\,\ref{fig:MS-impact} by at least the following two mechanisms not accounted for in our analysis: 1) Given the thermohaline convection, the pollution events should be observed less than 10 to 100\,Myr after their occurrence, and/or 2) the CZ is becoming so thin that even relatively small ($\sim10\,\mathrm{km}$) planetesimals may dissolve into the deeper radiative zone (below the CZ).
In both cases a fraction of the accreted rocky materials reaches the radiative zone and the enrichment of heavy elements in the CZ becomes smaller.

Let us now turn to the (unrealistic) case of the accretion of $0.03\,\Mstar$ metal-free gas onto MS stars. The final metallicity would be affected as a function of the mass of the convective zone, that is, $\Zs/{\Zini}=\MCZ/(\MCZ+0.03\,\Mstar)$. This implies that, using that approach, the retention of solids into planets would have changed the Sun's metallicity by half. Since this corresponds to a reduction of the CZ metallicity, this effect would not have been affected by fingering instabilities. Of course, this situation is not realistic: we must consider that the accretion of low-metallicity gas occurs when stars are on the pre-MS. This corresponds to a phase during which they have considerably larger convective zones, therefore greatly suppressing the dilution effect. 

\section{Stellar evolution with accretion} \label{sec:method}

In this section we describe how we model stellar evolution in the presence of accretion, and in particular how we account for a variable composition of the accreted gas. 
The method used is the same as \citetalias{Kunitomo+17}, but with some differences in the assumed accretion history, input physics, and in the temporal evolution of accreting materials' metallicity as described hereafter.

\subsection{Computation method} \label{sec:method-SE}
We use the one-dimensional stellar evolution code MESA version 6596 \citep{Paxton+11,Paxton+13,Paxton+15}.
Although we give a brief summary of the method below, we refer to the Paxton et al. papers and \citetalias{Kunitomo+17} for full details of the computational method.

We calculate the protostellar\footnote{We refer to the main accretion phase before stellar mass becomes comparable to its final value as the protostellar phase.
}, pre-MS and MS evolutions from a $0.01\,\Msun$, $1.5\,\Rsun$ seed including mass accretion.
Therefore, in this article the initial time (i.e., $t=0$) corresponds to the formation of a second Larson's core.
Although the initial entropy is important for the evolution of very-low-mass stars \citep[see][\citetalias{Kunitomo+17}]{Hosokawa+11}, we fix the initial condition, because we focus on stars more massive than $0.5\,\Msun$.
We stop calculations when the time reaches 10 Gyr or a star leaves its MS (its radius $R_{\star}>8\,\Rsun$), whichever occurs first.

Accretion flow injects not only materials but also entropy into a star.
Recent theoretical studies have shown that pre-MS evolution depends on the entropy of the accretion flow \citep[see][]{Hartmann+97,BCG09,BVC12,Baraffe+17,Hosokawa+11,Vorobyov+17,Jensen+Haugbolle18}.
We model the injected heat by the accretion per unit time, $L\sub{add}$, as a part of the gravitational energy of the accreting materials.
As in \citetalias{Kunitomo+17}, we introduce $\xi$, the fraction of accretion energy that is given to the star, such that
\begin{equation} \label{eq:xi}
L\sub{add}=\xi GM_{\star}\dot{M}/R_{\star}\,,
\end{equation}
where $\dot{M}$ is the mass accretion rate and $G$ the gravitational constant.
We assume a steady $\xi$ throughout the accretion for simplicity \citep[see however][]{BVC12}.
The rest of the gravitational energy, $(1-\xi)\,GM_{\star}\dot{M}/R_{\star}$, is assumed to be lost by radiation before the materials accrete onto the star.

In \citetalias{Kunitomo+17}, we found that {both the protostellar and} the pre-MS {evolutions depend} on how the accretion heat is distributed in the star. 
We adopt the following two models of the heat distribution.
{Our fiducial setting is to assume a uniform distribution in the entire star.}
Therefore, in this case the energy deposited in the star per unit mass and time is $\varepsilon\sub{add}= L\sub{add}/\Mstar$.
{We also consider the case of a non-uniform heat distribution, for which we introduce a dimensionless parameter, $m\sub{ke}$ defined such that the accretion heat is distributed only in the surface region that contains an amount of mass $m\sub{ke} M_\star$, and $\varepsilon\sub{add}$ increases linearly towards the stellar surface (see Eq.\,3 of \citetalias{Kunitomo+17}).
In the present article we show only one example of non-uniform heat distribution with $m\sub{ke}=0.1$.
}

\subsection{Mass accretion rate}
\label{sec:method-Mdot}

\begin{figure}[!t]
  \begin{center}
    \includegraphics[width=\hsize,keepaspectratio]{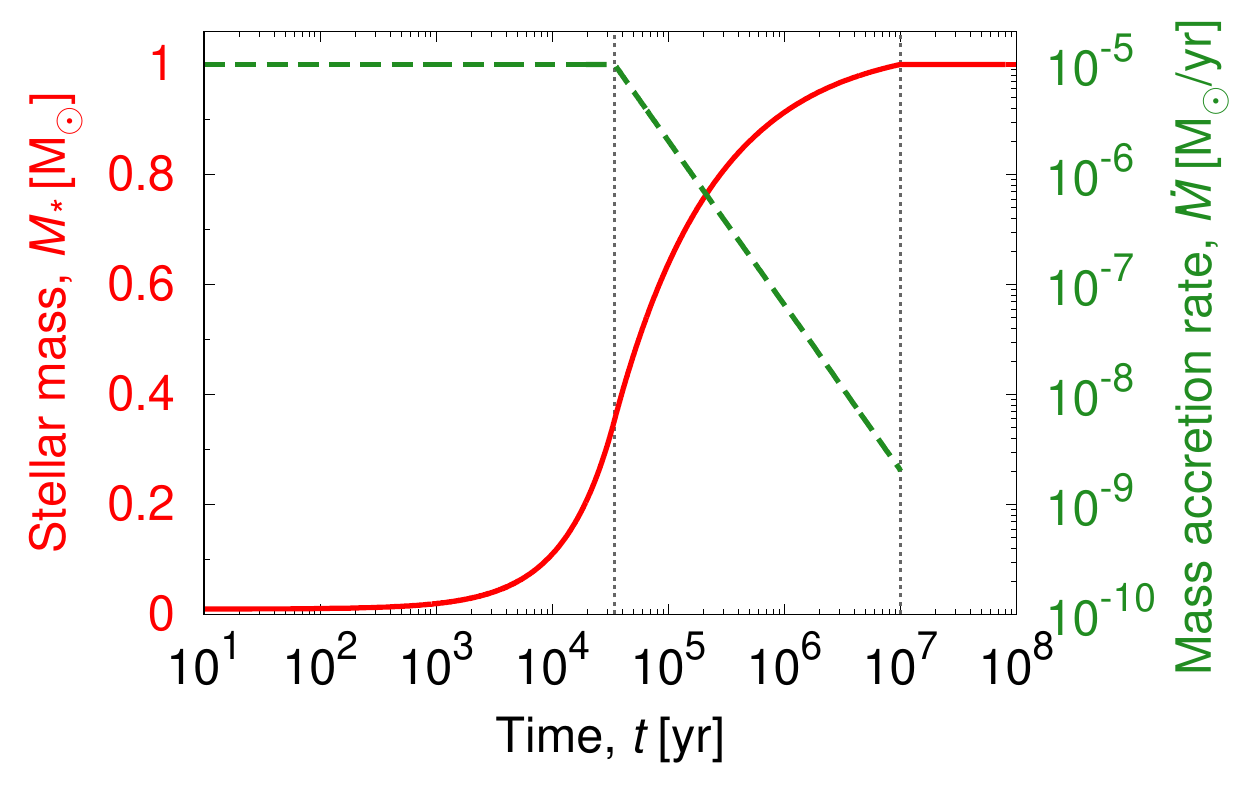}
        \caption{\small{{Temporal evolutions of stellar mass (solid line) and of the} accretion rate (dashed) in the case of $\Mfin=1\,\Msun$.
}}\label{fig:Hartmannt-M}
    \end{center}
\end{figure}

In \citetalias{Kunitomo+17}, we adopted a steady accretion of gas with $\dot{M}=10^{-5}\,\Msun/\mathrm{yr}$ for simplicity.
When studying the consequences of planet formation on the stellar surface composition, the important phases are after a few million years when the disk accretion is reduced, implying that it is important to {use} an accretion history that includes the decay in gas accretion rate with time. 
Following the observations in \citet{Hartmann+98}, we adopt the function of mass accretion rate with time shown in Fig.\,\ref{fig:Hartmannt-M}:
{it} is large and steady during the class I phase ($\dot{M}=10^{-5}\,{\Msun}/{\mathrm{yr}}\,(\Mfin/\Msun)$ until $\approx 3\times10^{4}\,\mathrm{yrs}$), 
whereas in the late phase it drops with time, $t$, as $\dot{M} \propto t^{-3/2}$.
The accretion rate is proportional to the final mass, $\Mfin$, and therefore the accretion stops at $10^{7}\,\mathrm{yr}$ for all cases, which corresponds to the upper limit of disk lifetimes, $t\sub{disk}$ \citep[see][]{Haisch+01,Hernandez+07,Mamajek09,Fedele+10,Ribas+14}.
For simplicity, we choose not to consider a correlation between stellar mass and the disk lifetime \citep[see, however, ][]{Hillenbrand+92,Hernandez+05,Yasui+14,Ribas+15}, and we do not include mass loss \citep[see however][]{Wood+05, Suzuki+13}.
The influence of varying the accretion history is described in Appendix\,\ref{sec:tdisk-zini}.

\subsection{Element diffusion}  \label{sec:method-diffusion}

Stellar composition evolves with time due to the following effects: 
thermonuclear reactions, convective mixing, {accretion,} element diffusion, and mass loss.
We include the first three in this article and neglect the others.
We adopt the Ledoux criterion for the convective stability.

Element diffusion (also referred to as microscopic diffusion or atomic diffusion) is thought to change surface compositions on long timescales ($\sim 1\,\mathrm{Gyr}$).
In general, it includes the effects of gravitational settling, {and} radiative levitation \citep[see, e.g.,][]{Chaboyer+01,Morel+Thevenin02,Dotter+17}.
For this work, we neglect any element diffusion process {and other extra-mixing processes} in radiative zones. 
This is because considering element diffusion with the standard MESA packages leads to a settling of heavy elements in MS high-mass stars which is incompatible with observations. 
Extra-mixing processes must prevent, at least partially, this settling. 
A proper consideration of diffusion (especially for stars beyond $1.2\,\Msun$) would be extremely valuable, but {since its efficiency is still a matter of debate \citep[see][]{Salaris+Cassisi17}, this is beyond the scope of this work. We note that diffusion and gravitational settling are thought to occur on $\sim\mathrm{Gyr}$ timescales, meaning that the results presented here should apply for pre-MS and early MS stars. For older stars we expect results to change quantitatively but not qualitatively. 
}

It should also be noted that our calculations do not include the effect of the fingering (thermohaline) convection (see Sect.\,\ref{sec:MSacc}). 
However, in cases where the {surface} metallicity is found to be larger than the interior metallicity, the expectation is simply that the star should rapidly homogenize so that its metallicity is uniform.

\subsection{Constraints on heat injection from \citetalias{Kunitomo+17}}  \label{sec:constraints}

We constrained the heat injection parameter, $\xi$, in \citetalias{Kunitomo+17} from the comparison of evolutionary tracks with observations of young clusters in the Hertzsprung-Russell (H-R) diagram.
We found that most stars would have been formed with $\xi\ga0.1$ if the mass fraction of deuterium (i.e., deuterium abundance) $X\sub{D}$ were found to be equal to $ 20\,\mathrm{ppm}$ in the clusters \citep{Hebrard+05,Steigman06,Linsky+06,Prantzos+07}.
Although this constraint depends on the uncertain deuterium abundances in the young clusters, we assume they are not very different from 20\,ppm and we adopt the constraint $\xi\ga0.1$ in this article.

This constraint also appears to match the ``hot'' or ``hybrid'' accretion scenario favored by \cite{Vorobyov+17} on the basis of the peak luminosities of FU-Orionis outbursts. 
We however show hereafter the $\xi=0$ cases to show the global trends with $\xi$.

Although \citetalias{Kunitomo+17} included element diffusion and used a different accretion history (namely, steady accretion), the constraints obtained in that article still apply: the thermal evolutions of pre-MS and MS stars in the H-R diagram are not sensitive to accretion history \citep[see][\citetalias{Kunitomo+17}]{Hosokawa+11} and element diffusion.

\subsection{Input parameters}  \label{sec:param}

We use the following seven parameters in the stellar evolution calculations (with fiducial values in parenthesis): 
$\xi (0.1)$,  $m\sub{ke}$ (uniform), $X\sub{D} (28\,\mathrm{ppm})$, $\Mfin (1\,\Msun)$, accretion history ($t\sub{disk}=10\,\mathrm{Myr})$, $Z\sub{ini} (0.01966\equiv\Zsun)$, and initial condition $(0.01\,\Msun, 1.5\,\Rsun)$.
Another parameter is the mass of zero- (or low-) metallicity gas accreted in the final stages of stellar accretion: $\dMO\ (0.03\,\Mfin)$ (see Sect.\,\ref{app:Mrock-extra}).

As described in the previous section, we set our fiducial $\xi$ value to 0.1 with uniform heat distribution.
We explore different heat injection hypotheses by varying $\xi$ and $m\sub{ke}$ in Sect.\,\ref{sec:int-evols}.
We found in \citetalias{Kunitomo+17} that the protostellar evolution depends on $X\sub{D}$.
Since we focus on {relatively old ($\ga1$\,Gyr) MS} stars including the Sun, we set our fiducial $X\sub{D}$ to the protosolar nebula value, 28\,ppm \citep[the observational uncertainty is $\pm2.8\,\mathrm{ppm}$,][]{Asplund+09}, rather than the present-day local ISM value, 20\,ppm. We also explore solutions with  $X\sub{D}$ varying from 0 to 40\,ppm.
{Deuterium abundance ($X\sub{D}$) in the accreting materials} is kept constant throughout the {whole accretion phase}.

   \begin{table}[!tb]
      \caption[]{Results of the $\chi^2$ test to reproduce the observed solar values without effects of the element diffusion.}
         \label{tab:solar} \small
         \begin{tabular}{p{1.5cm}lp{2.6cm}}
            \hline
            \hline
            \noalign{\smallskip}
            & \multicolumn{2}{l}{Converged input parameters} \\
            \noalign{\smallskip}
            \hline
            \noalign{\smallskip}
            $X_{\mathrm{ini}}$                  & 0.70048186 & \\
            $Y_{\mathrm{ini}}$                  & 0.27985523 & \\
            $Z_{\mathrm{ini}}$                  & 0.01966291 & \\
            $\alpha\sub{MLT}$                   & 1.81743512 & \\
            $f_{\mathrm{ov}}$                   & 0.01026355  & \\
            \noalign{\smallskip}
            \hline
            \noalign{\smallskip}
             & Target values & Converged values at solar age\\
            \noalign{\smallskip}
            \hline
            \noalign{\smallskip}
            $(Z/X)_{\mathrm{surf}}$                     & $0.02313\tablefootmark{a}$ & 0.0280705368   \\
            $Y_{\mathrm{surf}}$                         & $0.2485\pm0.0035\tablefootmark{b}$ & 0.2798552573 \\
            $R_{\mathrm{CZ}}/\Rsun$                             & $0.713\pm0.001\tablefootmark{c}$ & 0.7227136520   \\
            $R_{\star}$                                         & $6.9598 \times 10^{10}\,{\mathrm{cm}}\tablefootmark{c}$  & $0.9996930658\,\Rsun$  \\
            $L_{\star}$                                         & $3.8418 \times 10^{33}\,{\mathrm{erg/s}}\tablefootmark{c}$ & $1.000011699\,\Lsun$  \\
            $\teff\,[{\mathrm{K}} ]$      & $5776.13157\tablefootmark{d}$ & $5776.93324$   \\
            \noalign{\smallskip}
            \hline
            \noalign{\smallskip}
         \end{tabular} 
         \tablefoot{
         $^{(d)}$ Calculated with $L_\star$, $R_\star$ and Stefan-Boltzmann law.
         {\bf References.}
         $^{(a)}$ \citet{GS98};
         $^{(b)}$ \citet{Basu+Antia04};
         $^{(c)}$ \citet{Bahcall+05}. }
        \normalsize
            \end{table}

\begin{figure}[!h]
  \begin{center}
    \includegraphics[width=\hsize,keepaspectratio]{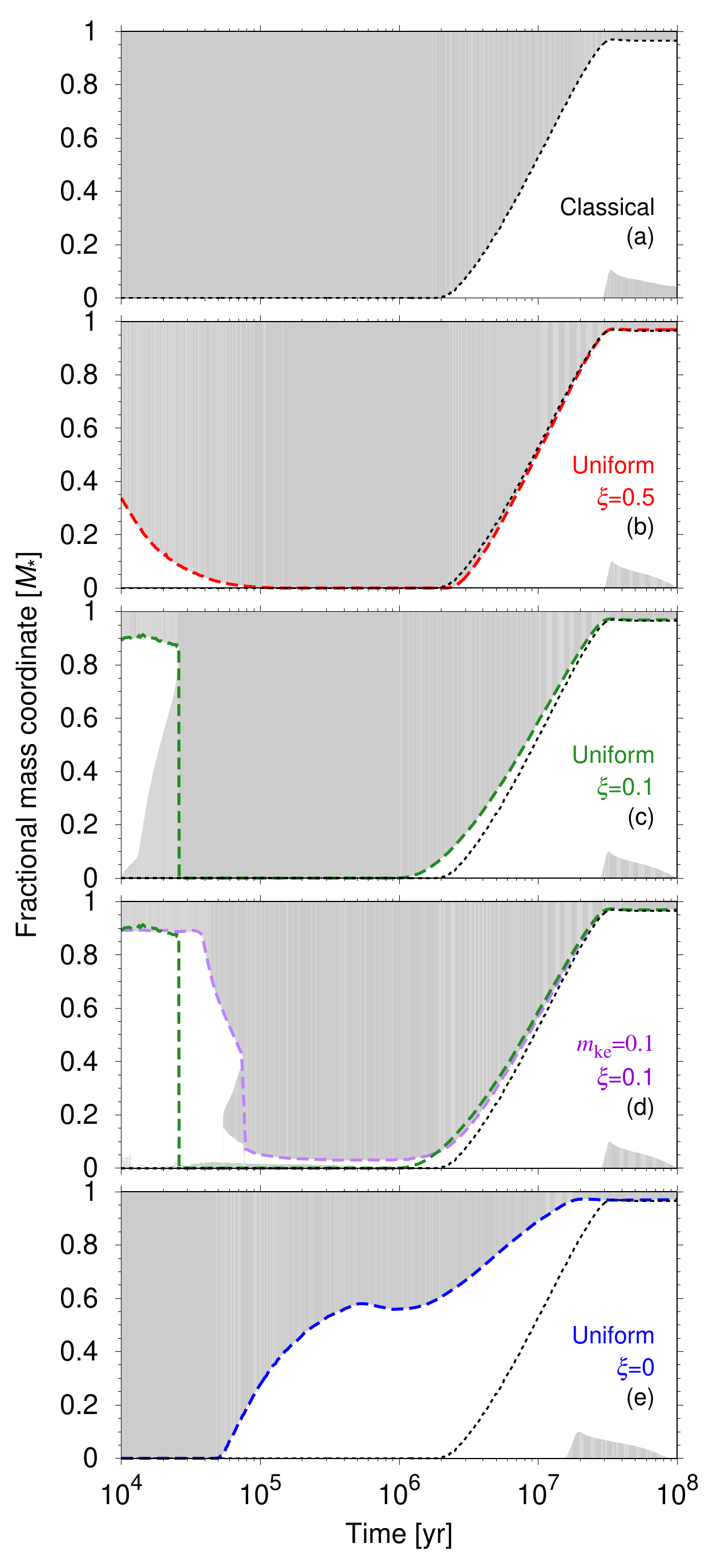}
        \caption{\small{
        Temporal evolution of the internal structure of stars with $\Mfin=1\,\Msun$ and $X\sub{D}=28\,\mathrm{ppm}$ in the cases of  (a) a classical, non-accreting model and  (b)--(e) accreting models.
        The vertical axes are the mass coordinates normalized by stellar mass.
        The shaded regions indicate the convective regions and the lines show the base of the surface CZ.
        In panels (b), (c), and (e), $\xi=0.5$, 0.1, and 0, respectively, with uniform heat distribution.
        In panel (d), the non-uniform heat distribution is assumed: $\xi=0.1$ and $m\sub{ke}=0.1$.
        For comparison, each panel includes the dotted line in panel (a).
        In addition, the CZ base of (c) is also illustrated in panel (d) by the green dashed line.
        }}\label{fig:conv-multi}
    \end{center}
\end{figure}

We use the parameters of composition and convection listed in Table\,\ref{tab:solar} throughout this article.
In \citetalias{Kunitomo+17}, we performed a $\chi^{2}$ test to explore the best settings to reproduce the spectroscopic and helioseismic observations of the present-day Sun.
Since we do not include the element diffusion in this article, we have again done the test and obtained the results listed in the table.
We note that the different input parameters in the table do not significantly affect the evolution of pre-MS and MS stars
{except for $\Zini$, which is discussed in Appendix\,\ref{sec:tdisk-zini}.}

\subsection{Protostellar and pre-MS evolution of stellar internal structure} \label{sec:int-evols}

As a first step, we discuss the results of accreting stellar models adopting different input settings, namely by changing the heat injection parameter (i.e.,  $\xi$ and  $m\sub{ke}$), the mass fraction of abundance $X\sub{D}$ and the final mass of stars $\Mfin$.
We emphasize that in this subsection we do not consider the effect of planet formation and therefore the composition of accreting materials are kept constant throughout the accretion, {unlike the following sections}.

Figure\,\ref{fig:conv-multi} shows the consequences of varying the heat injection parameter between $\xi=0$ and 0.5 {for $1\,\Msun$ stars}.
The qualitative behaviors are the same: 
{in} an initial phase of less than $10^5$\,yrs in which the star has accreted less than half of its final mass, the interior is largely convective but {radiative} regions can exist, depending on how the accretion takes place (i.e., depending on the parameters $\xi$ and $\mke$). 
After that, the star becomes (almost) entirely convective until after $10^6$\,yrs or so (for classical and $\xi\geq0.1$ models), when a radiative core develops and grows. 
We note that $\xi=0$ is a special case,
{because} the star is so compact that it starts fully convective and the radiative zone appears and grows before $10^5$ years, much earlier than for $\xi\wig{>}0.1$. 
The timing therefore strongly depends on the value of $\xi$ \citep[see also][]{BC10}. 
For example, the time when a {stable} radiative core {forms} is 1.8, 2.2, 0.9, and {0.05\,Myr}
and the time at which $\MCZ=0.1\,\Msun$ is 24, 24, 22 and 10\,Myr for a non-accreting case, $\xi=0.5$, $\xi=0.1$ and $\xi=0$, respectively.

The underlying physics of the {protostellar and} pre-MS evolution of stellar internal structure responsible for the evolution in Fig.\,\ref{fig:conv-multi} is discussed in Appendix\,\ref{sec:physics}. 
In short, the stellar internal structure is controlled by three effects: heat injection by accretion, opacity, and delivery of fresh deuterium in the stellar interior.

\begin{figure}[!t]
  \begin{center}
    \includegraphics[width=\hsize,keepaspectratio]{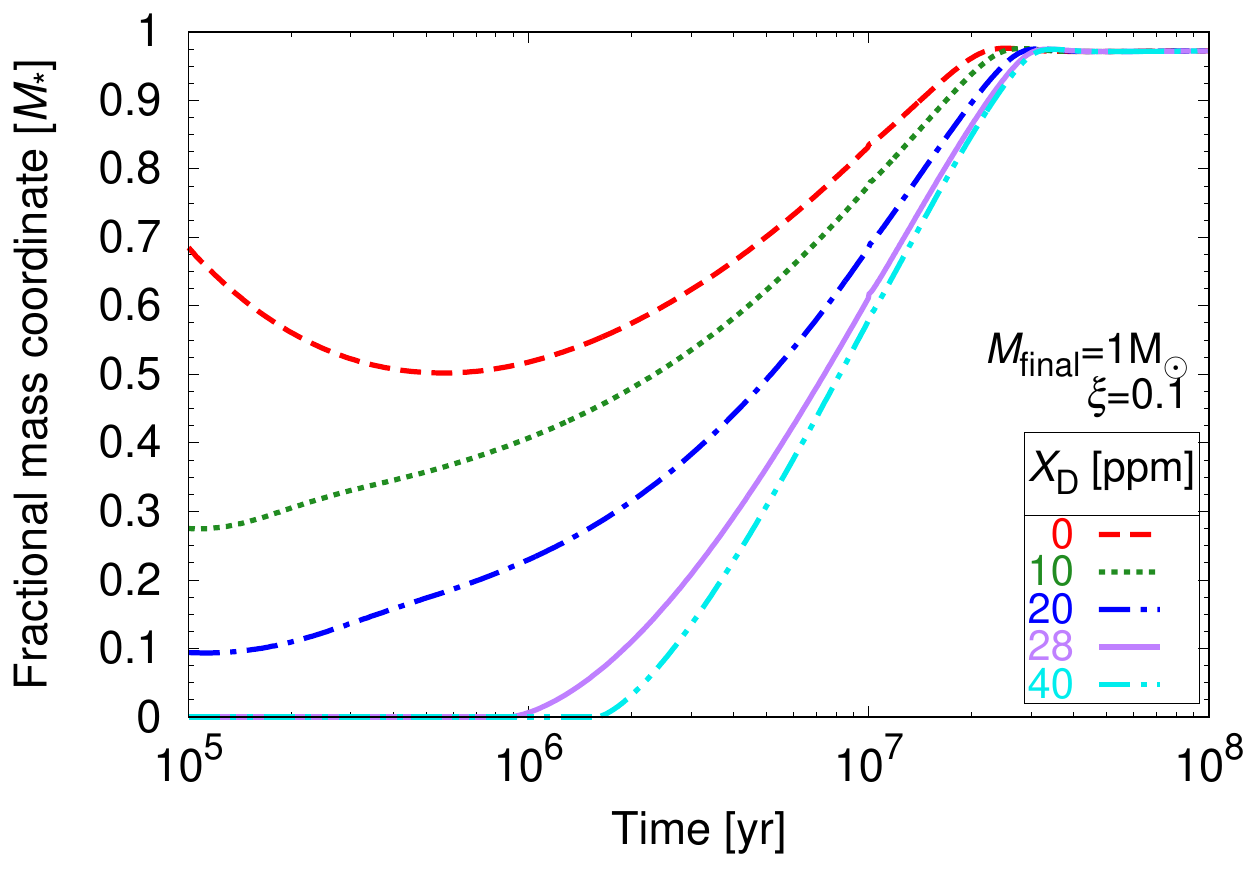}
        \caption{\small{
        Temporal evolutions of the location of the surface CZ base normalized by stellar mass as in \figref{fig:conv-multi} with $X\sub{D}=0, 10, 20, 28$ and 40\,ppm from top to bottom.
         In all cases $\Mfin=1\,\Msun$ and $\xi=0.1$ with the uniform heat distribution.
        }}\label{fig:D-conv}
    \end{center}
\end{figure}

Figure\,\ref{fig:D-conv} shows the evolution of the location of the radiative/convective interface for different $X\sub{D}$ values ranging from 0 to $40\,\mathrm{ppm}$ (the cosmic primordial value) {with $\xi=0.1$}.
The protosolar value is $28\,\mathrm{ppm}$ and the value for the current local interstellar medium is $\approx20\,\mathrm{ppm}$.
Although variations of $\xi$ have a quantitatively larger impact, variations in $X\sub{D}$ also affect the internal structure evolution: the stars with a lower $X\sub{D}$ develop a radiative core more rapidly.
This is because such stars have a smaller radius (see Fig.\,3 of \citetalias{Kunitomo+17}), 
a higher temperature ($T\propto M_{\star}/R_{\star}$ from the virial theorem),
and then a smaller opacity {($\kappa\propto T^{-3.5}$, using Kramers' law)}.

The importance of the choice of deuterium abundance is illustrated by the fact that for cold accretion ($\xi=0$), if $X\sub{D}=20\,\mathrm{ppm}$, $\MCZ=0.1\,\Msun$ only after a few million years \citep[see also Fig.\,4 of ][]{BC10}, while for $X\sub{D}=28\,\mathrm{ppm}$, reaching this point takes $22\,\mathrm{Myr}$ (see \figref{fig:conv-multi}c).

\begin{figure}[!t]
  \begin{center}
    \includegraphics[width=\hsize,keepaspectratio]{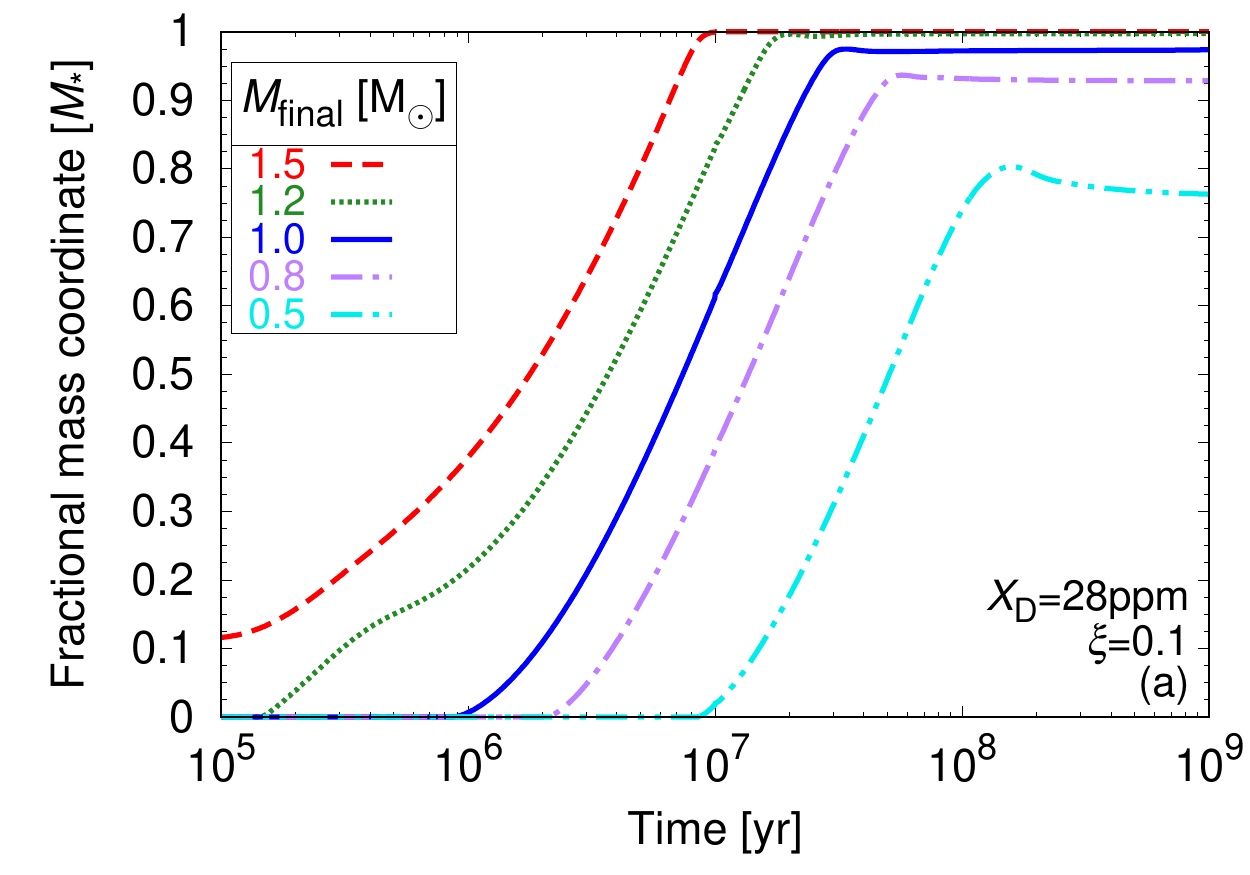}
    \includegraphics[width=\hsize,keepaspectratio]{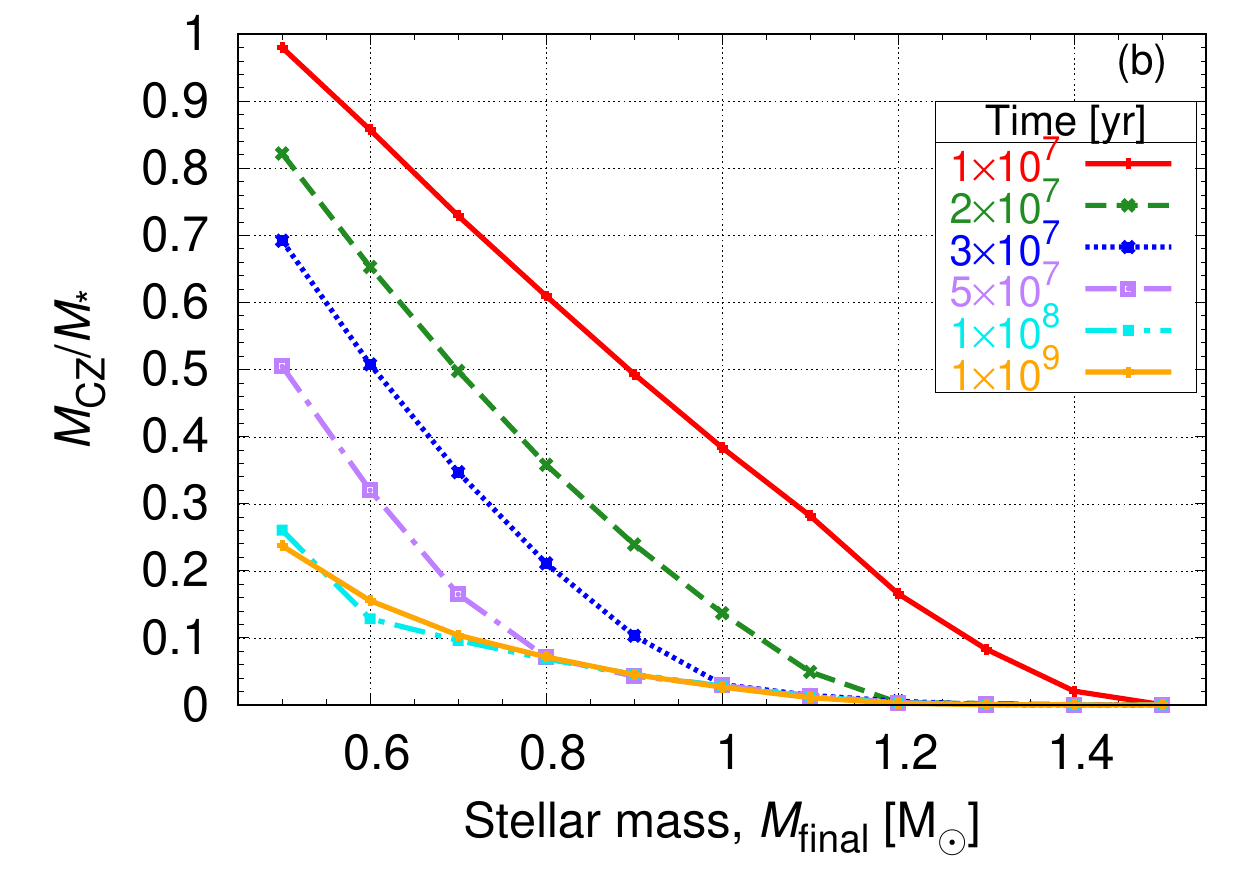}
        \caption{\small{
        {\textit{Top panel.}} 
        Temporal evolution of the surface CZ base for the following final stellar masses: $\Mfin= 1.5, 1.2, 1, 0.8$ and $0.5\,\Msun$ from top to bottom. 
        In all cases, accretion stops at $10^7$ years, and $\xi=0.1$, $X\sub{D}=28\,{\mathrm{ppm}}$ and a uniform heat distribution are assumed. 
        {\textit{Bottom panel.}}
        Mass fraction of the CZ as a function of stellar mass.
        The lines from top to bottom denote the temporal evolution from $10^{7}$ to $10^{9}$ yrs.
        }}\label{fig:Mfin-conv}
    \end{center}
\end{figure}

Figure\,\ref{fig:Mfin-conv}a shows how the convective zone mass changes as a function of age and for a range of final masses $\Mfin=0.5$--$1.5\,\Msun$.
We note that $\dot{M}\propto\Mfin$ and then the accretion stops at $10^{7}\,\mathrm{yrs}$ in all cases (see Sect.\,\ref{sec:Za}).
More massive stars have a thinner surface CZ because a larger mass corresponds to a hotter interior.
Figure\,\ref{fig:Mfin-conv}b shows the temporal evolution of $\MCZ$ as a function of stellar mass from $10^{7}$ to $3\times10^{9}\,\mathrm{yrs}$.
The lower-mass stars have a longer timescale for the shrinkage of the surface CZ due to a longer Kelvin-Helmholtz (K-H) timescale.

\subsection{Composition of accreting materials} 
\label{sec:Za}

The accretion of condensates (i.e., heavy elements), and therefore the evolution of the metallicity of the gas accreted by the star, $\Za$, is controlled by the growth of dust grains and the gas drag by the disk which cause them to quickly drift towards the star \citep[see, e.g.,][]{Garaud07}, and, when planetesimals and protoplanets form, by their filtering by these larger bodies \citep{Guillot+14}. In the early phase, before grains have the time to grow, we expect the star to accrete material that has the composition of the molecular cloud core, $\Zini$. When grains grow but few large non-drifting planetesimals have been formed, we expect $\Za$ to increase \citep{Garaud07, Tsukamoto+17}. When planetesimals and planets form, $\Za$ should decrease, and eventually become very small when large-enough protoplanets (say from a few to a few tens of Earth masses) form and filter any incoming dust \citep{Guillot+14, Morbidelli+16}. A large variety of scenarios may therefore be considered.

We choose a very simple scenario that appears to best capture the essence of the problem:
in the early phase $\Za(t<\tp)=\Zini$, and then when large-enough planets have formed, $\Za(t\ge\tp)=0$. (In Sect.\,\ref{sec:discussion-solar}, we also consider the cases in which ${\Zini}>{\Za}(t\ge\tp)>0$.)

One could of course argue that in the no-planet case, grains could be lost rapidly, leaving a tail of low-metallicity accretion that could mimic the case with planets. This is what happens for example in the model by \citet{Garaud07} and other similar grain-growth and drift models. However, on one hand, this efficient loss of grains onto the central star is incompatible with observations \cite[see][]{Dullemond+Dominik05,Takeuchi+Lin05, Brauer+07}; on the other hand, for systems with planets, the additional reduction in metallicity is probably much larger.

One may also ask why, with our simple model, we do not consider the possibility of this initial increase in metallicity. 
{While} this may be of interest for some applications, this increase in metallicity is likely to be erased rapidly for two reasons: (1) because it occurs relatively early, when the protostar is still largely convective, and (2) because even in the case of the radiative region, as discussed previously, the top-heavy situation is likely to be unstable to double-diffusive instabilities. 

Of course, our approximation that the metallicity is constant until it becomes zero is extreme and will lead to an overestimate of the magnitude of the effect of planet formation on stellar composition. In view of the other approximations, in particular on the assumed mass that is taken by planets, we believe that this is minor. Our scenario is to be considered as a guide to estimate the magnitude of the effects of planet formation rather than the basis of a quantitative, predictive model.

\subsection{Stellar surface metallicity calculation} \label{sec:model_comp}

Finally, we describe how we calculate temporal evolution of $\Zs$.
Ideally, by performing stellar evolution calculations in which $\Za$ changes with time (``full simulation''), we can obtain accurate solutions for the evolution of $\Zs$.
The full simulation takes time, however.
Instead, here we solve the $\Zs$ evolution in the following way (``approximate solution''):
{first} we perform a stellar evolution calculation with a constant $Z\sub{acc}=\Zini$.
Using the temporal evolution of $M\sub{CZ}$ and the accreted mass in each time step, {$\delta M(t_{i})$,} we can estimate the evolution of $\Zs$ with a given evolutionary model of $Z\sub{acc}$ as
\begin{equation} \label{eq:simpleZs}
\Zs(t_{i})=
\frac
{\MCZ(t_{i-1})\,\Zs(t_{i-1}) + \delta M(t_{i})\,\Za(t_{i}) }
{\MCZ(t_{i-1})+\delta M(t_{i})}
\ ,
\end{equation}
where $t_{i}$ is the time at $i$-th step.
Our post-processing method is fast and therefore enables us to simulate a large variety of accreting models.

We note that in the full simulation models, the decrease in the metallicity $\Za$ (due to the planet formation) is compensated for by an increase in the mass fraction of $^{1}\mathrm{H}$ and $^{4}\mathrm{He}$ in the accreting materials.
We recall that $\sum X_{i}=1$, where $X_{i}$ is the mass fraction of each element $i$ in mass.
Therefore, we redistribute the missing metallicity $\Delta Z$ by adding $0.7\Delta Z$ to $X_{^{1}\mathrm{H}}$ and $0.3\Delta Z$ to $X_{^{4}\mathrm{He}}$.

\begin{figure}[!t]
  \begin{center}
    \includegraphics[width=\hsize,keepaspectratio]{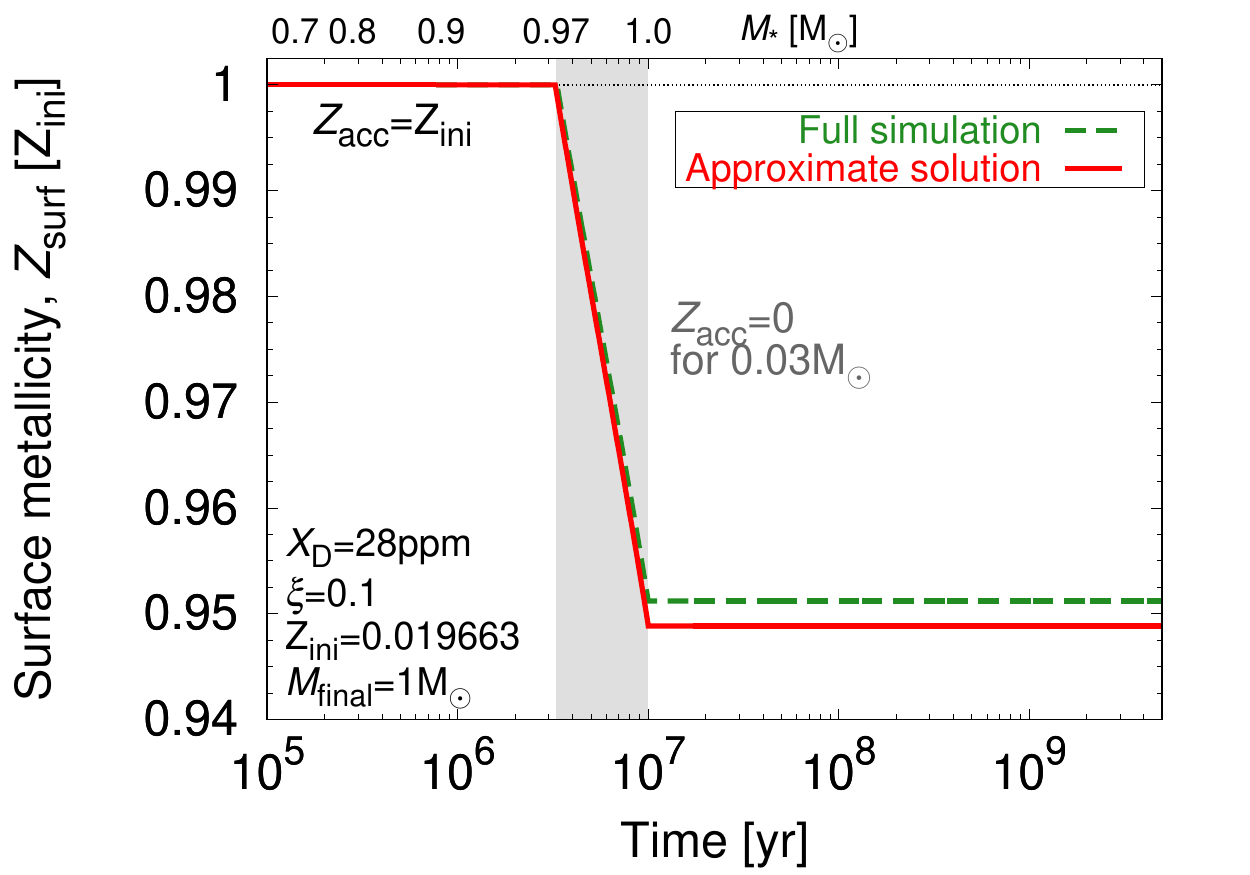}
        \caption{\small{
        Temporal evolution of {stellar surface metallicity}, $\Zs$, normalized {to} initial metallicity, $\Zini$, with $\Mfin=1\,\Msun$, $\xi=0.1$ and $X\sub{D}=28\,\mathrm{ppm}$.
        At first $\Za=\Zini$, whereas $\Za=0$ during the last $0.03\,\Msun$ (shaded).
        The approximate solution of $\Zs$ (red, solid line) well matches the full simulation (green, dashed line).
        }}\label{fig:comp-Zs}
    \end{center}
\end{figure}

Let us verify the approximate solution.
Figure\,\ref{fig:comp-Zs} shows the comparison between the evolutions of $\Zs$ using the full simulation and the approximate solution.
We adopt $\Mfin=1\,\Msun$ and the fiducial settings (i.e., the evolution of $\Za$ described in Sect.\,\ref{sec:method-Mdot}, $\xi=0.1$ with the uniform heat distribution, and $X\sub{D}=28\,\mathrm{ppm}$).
We find that the two are almost the same: in both cases $\Zs$ is decreased by $\approx5\,\%$ due to the metal-free gas accretion in the late phases.
This fact validates the approximate solution and therefore we use it in Sect.\,\ref{sec:discussion}.

{We note that $\log(\Zs/\Zini)$ and $\dfeh$ are related through the relation
\begin{equation}
\dfeh \equiv \log({\Zs}/{\Zini}) - \log(X\sub{surf}/\mathrm{X_{ini}})\,.
\end{equation}
Differences between these two quantities can be neglected since $|\log(X\sub{surf}/\mathrm{X_{ini}})| < 0.03$ even in the cases where $\log(\Zs/\Zini) = \pm0.6$. In this work we use $\log(\Zs/\Zini)$ and $\dfeh$ interchangeably.
}

\section{Stellar composition anomalies resulting from planet formation} \label{sec:discussion}

In this section we investigate the consequences of planet formation on the stellar surface composition using the results in Sect.\,\ref{sec:int-evols}.
We then compare our predictions to the observations of the Hyades cluster and the $\lambda$ Boo stars.

\subsection{Dilution of stellar surface composition} \label{sec:Zsurf}

\begin{figure}[!t]
  \begin{center}
    \includegraphics[width=\hsize,keepaspectratio]{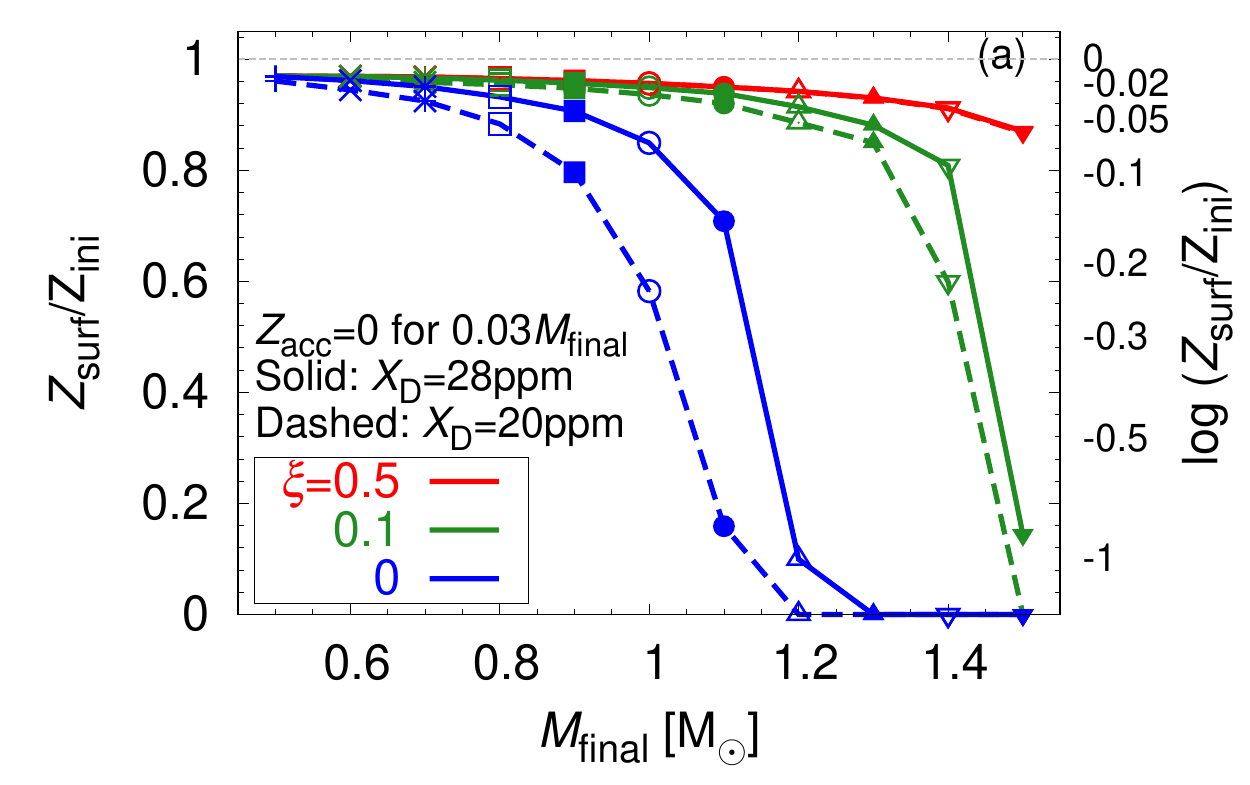}
    \includegraphics[width=\hsize,keepaspectratio]{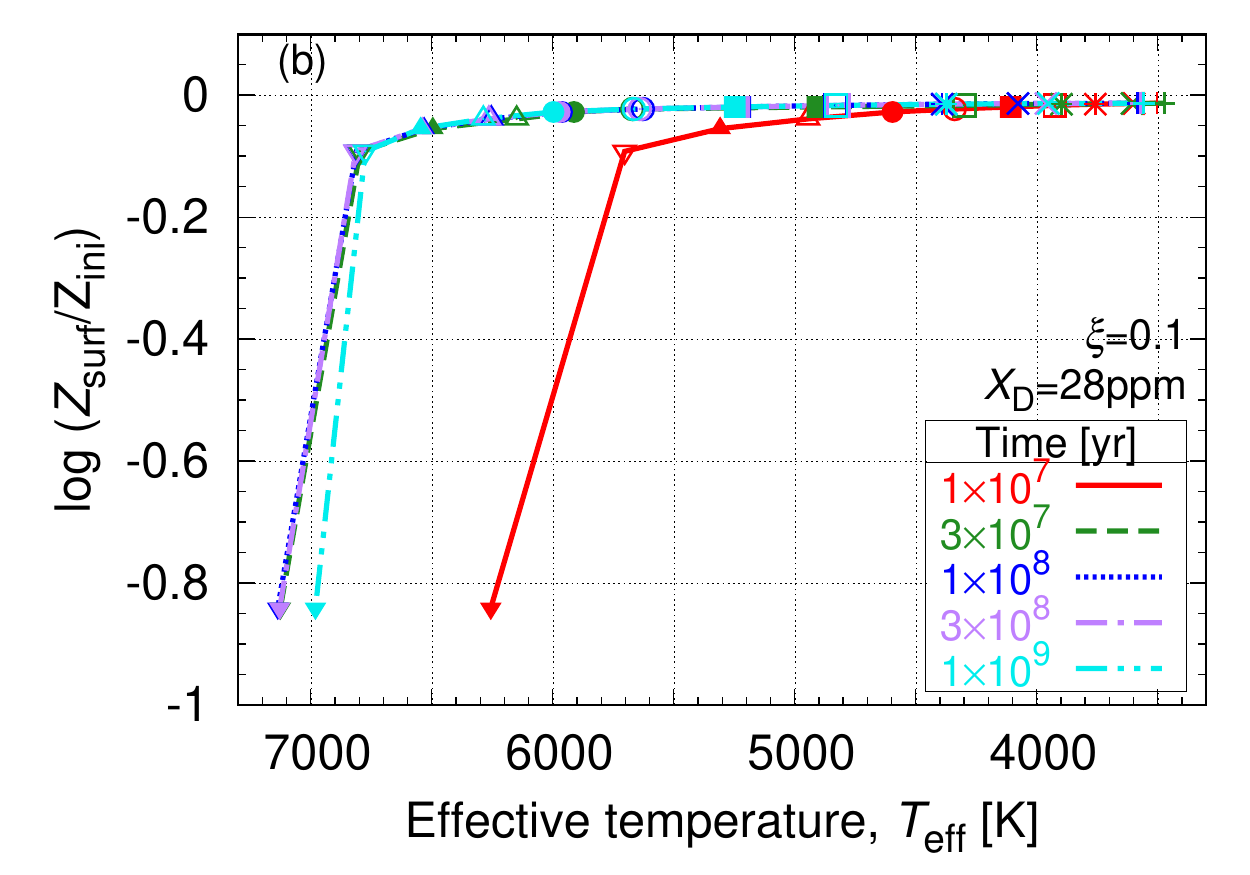}
        \caption{\small{
        \textit{Top panel.}
        Stellar surface metallicity (normalized by the initial metallicity) after a metal-free accretion during the last $0.03\,\Mfin$ as a function of stellar mass.
        The solid and dashed lines denote the cases with $X\sub{D}=28$ and 20\,ppm, respectively, and $\xi=0.5, 0.1$ and 0 from top to bottom.
        In the $\xi=0.5$ case, both lines are overlapped.
        \textit{Bottom panel.}
        Surface metallicity as a function of the stellar effective temperature, $\teff$, in the case with $\xi=0.1$ and $X\sub{D}=28\,\mathrm{ppm}$ (the green dashed line in the top panel).
        The same points are used as in the top panel.
        }}\label{fig:M-Zsurf}
    \end{center}
\end{figure}

We use our ``approximate solution'' to calculate the evolution of the stellar surface metallicity $\Zs$ resulting from 
{the ingestion of metal-free materials (Sect.\,\ref{sec:Za})}.
Figure\,\ref{fig:M-Zsurf}a shows the value of $\Zs$ for stars in their MS phase and with $\Mfin=0.5$--$1.5\,\Msun$, $\xi=0$--0.5 and $X\sub{D}=20$ and 28\,ppm.
{(For simplicity, from this section onward, we present only results obtained with a uniform heat distribution.)}
We note that because we do not include element diffusion, once on the MS, $\Zs$ is constant with time.
The magnitude of the change in $\Zs$ due to planet formation depends on the three parameters $\Mfin$, $\xi,$ and $X\sub{D}$.
The resultant $\Zs$ is lower with a higher $\Mfin$, a lower $\xi$ (which is predominant) and a lower $X\sub{D}$.
This is because the $\MCZ$ evolution is a function of the three parameters, as shown in Figs.\,\ref{fig:conv-multi}--\ref{fig:Mfin-conv}: 
{in} the late accretion phase, higher-mass stars formed with lower $\xi$ and $X\sub{D}$ have a thinner CZ. 
In that case, for the same amount of condensates retained into planets, the stellar surface composition is affected more significantly.

Since the mass is not generally available as observable,
we show $\Zs$ as a function of the stellar effective temperature, $\teff$, in Fig.\,\ref{fig:M-Zsurf}b in the cases with $\xi=0.1$ and $X\sub{D}=28\,\mathrm{ppm}$.
Although $\Zs$ is constant with time after the accretion phase, the $\Zs$ -- $\teff$ relation evolves with time because of the evolution of $\teff$ on the pre-MS.

The solid lines in Fig.\,\ref{fig:Teff-Zsurf-planetacc} show the consequences of the dilution on $\Zs$ with $\xi=0.5, 0.1$ and 0 and $X\sub{D}=28\,\mathrm{ppm}$ in the {$\teff$ -- $\dfeh$ diagram}. We find that changes in $\Zs$ due to planet formation are small in the (standard) $\xi=0.5$ case. However, for cold accretion ($\xi=0.1$ and 0), stars with effective temperatures above $6000$ to $7000\,\mathrm{K}$ can show an important reduction in their metallicity due to dilution.

As described in Sect.\,\ref{app:Mrock}, here we adopt a $0.03\,\Mfin$ zero-metal accretion as a planet formation model, corresponding to an upper limit in the planetary systems with giant planets.
If there is variability in planet formation, a scatter in $\log (\Zs/\Zini)$ should be created between 0 (i.e., no planet formation) and the solid line in Fig.\,\ref{fig:Teff-Zsurf-planetacc} (see also Table\,\ref{tab:feh}).
For example, in our fiducial setting (i.e., $\xi=0.1$ and $X\sub{D}=28\,\mathrm{ppm}$), we predict a maximal spread of [Fe/H] of 0.1\,dex at around $\teff=6800\,\mathrm{K}$ in a cluster.
On the other hand, for lower $\teff\,(\la6000\,\mathrm{K})$, a much smaller scatter ($<0.03\,\mathrm{dex}$) can be created by the dilution.

\subsection{Pollution by planetesimal engulfment during the MS} \label{sec:MS}

Figure\,\ref{fig:Teff-Zsurf-planetacc} also shows the consequences of the engulfment of heavy elements onto MS stars (i.e., the pollution effect) {assuming that $\dM\sub{acc}=1$ or $0.01\,\Mearth$ and $\Za=1$ in Eq.\,\eqref{eq:Zs-MS}}.
We consider that the stars experienced metal-free accretion during their pre-MS phase, meaning that their surface composition was diluted before they began being impacted by planetesimals.

We find that for high-mass stars $(\ga6500\,\mathrm{K})$, the engulfment of heavy elements can increase $\Zs$ significantly $(\ga0.1\,\mathrm{dex})$.
In the $\xi\ga0.1$ cases, even if $\Zs$ is much lower than $\Zini$ due to the dilution during the pre-MS phase, the engulfment can completely overcome the initial decrease in $\Zs$ due to the dilution effect. In the $\xi=0$ cases, however, planetesimal engulfment cannot compensate for the strong dilution.

We note that because of  fingering (thermohaline) convection \citep{Vauclair04,Garaud11,Theado+Vauclair12,Deal+15}, we expect any increase in $\Zs$ above its initial value $\Zini$ to be quickly erased (see Sect.\,\ref{sec:MSacc}).
Furthermore, although the effect of gravitational settling is still a matter of debate and is not included in this article, it has the potential to change the stellar surface metallicity with a long timescale ($\sim1\,\mathrm{Gyr}$).

There is however a case in which we may be observing the signature of pollution.
In the pre-MS cluster $\gamma$ Vel, the star ``2MASS J08095427--4721419'' is more metal-rich compared to other members by $\dfeh\approx0.13\,\mathrm{dex}$ \citep{Spina+14a}.
Since the cluster is young ($\sim15\,\mathrm{Myr}$), fingering convection may not have had time to erase any pollution effect.
It is interesting to note that orbital instabilities in planetary systems are expected to generally occur soon (with $\sim10\,\mathrm{Myr}$ or so) after disk dispersal \citep[see][]{Iwasaki+01, Kominami+Ida02}. Given this scenario, pre-MS stars could potentially have increased metallicities due to a recent pollution event. On the other hand, such events would in most cases be ancient in MS stars meaning that their signatures would have been erased by fingering convection.

\subsection{Comparison with observations of the Hyades cluster} \label{sec:obs}

\begin{figure}[!t]
  \begin{center} 
  \includegraphics[width=\hsize,keepaspectratio]{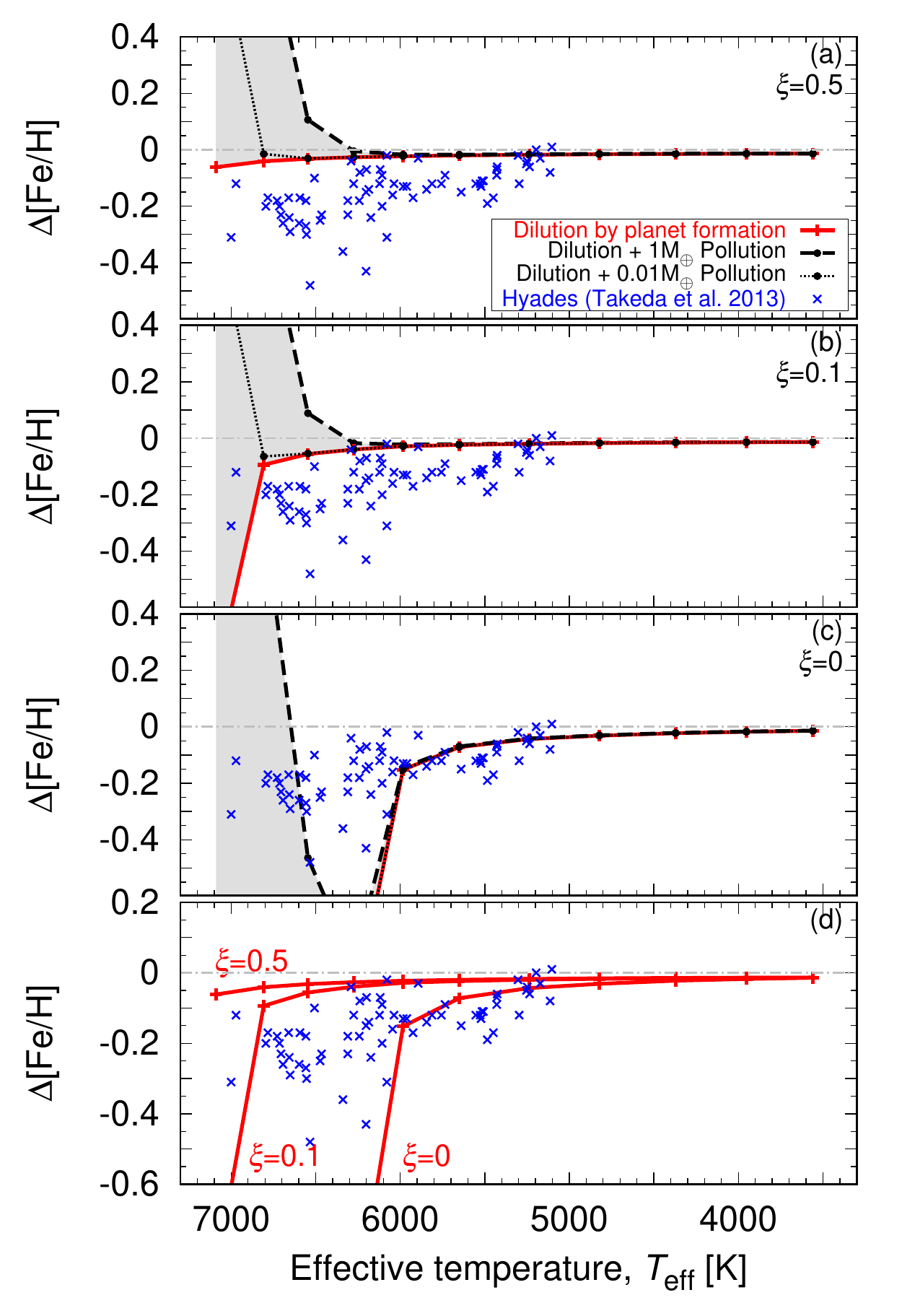}
        \caption{
\small{
        Stellar surface metallicity variation $\dfeh$ ($\approx \log(\Zs/\Zini)$; see text) resulting from planet formation as a function of effective temperature for stars in a cluster (same age, same metallicity). 
        Three scenarios are considered: the dilution of the stellar surface due to the late accretion of $0.03\,\Mfin$ of metal-free gas (solid red lines), the same dilution plus a pollution by the engulfment of $0.01\,\Mearth$ (dotted lines) or $1\,\Mearth$ (dashed lines) of planetesimals, respectively (see text). Observations of the 0.6-Gyr-old Hyades cluster \citep{Takeda+13} are shown as blue crosses.
        All model values are evaluated at an age of 0.6\,Gyr with $X\sub{D}=28\,\mathrm{ppm}$.
        {The first three panels show model $\xi$ values of} 0.5 (a), 0.1 (b) and 0 (c).
        The bottom panel (d) illustrates the possible range of $\Zs$ created only {by the dilution scenario as a function of $\xi$}.
        }}\label{fig:Teff-Zsurf-planetacc}
    \end{center}
\end{figure}

In Fig.\,\ref{fig:Teff-Zsurf-planetacc}, we compare our predictions with observations of the Hyades cluster.
We assume that cool stars in the sample (with $\mathrm{[Fe/H]\sim0.25}$) are representative of the primordial, non-diluted metallicity and the primordial composition of the cluster members were uniform.
We therefore plot $\mathrm{\Delta[Fe/H] = [Fe/H] - 0.25}$ as a function of effective temperature. Given our assumptions, we note that the results for the dilution scenario as plotted here are insensitive to the assumed metallicity (see Appendix\,\ref{sec:tdisk-zini}).
The observed data from \citet{Takeda+13} show a trend in the $\teff$ -- [Fe/H] plane: 
{hotter} stars have a lower metallicity ($\sim0.3$\,dex over 2000\,K from 5000 to 7000\,K) with a significant (up to 0.4\,dex) scatter.
First we stress that it is difficult to precisely determine the metallicity of rapid-rotating hot stars and therefore this $\teff$ -- [Fe/H] trend may be created by uncertainties in atmospheric models \citep[see][]{Takeda+13}. We however want to explore the possibility that the trend and/or the scatter are caused by planet formation.

   \begin{table}[!tb]
      \caption[]{Expected scatter in [Fe/H] in dex induced by the dilution.}
         \label{tab:feh} \small
         \begin{tabular}{p{0.4cm}|p{0.9cm}p{0.9cm}p{0.9cm}p{0.9cm}p{0.9cm}p{0.9cm}}
            \hline
            \hline
            \noalign{\smallskip}
            $\xi$ &  \multicolumn{6}{c}{$\teff$\,[K]}\\
            & 7000 & 6500 & 6000 & 5500 & 5000 & 4000\\
            \noalign{\smallskip}
            \hline
            \noalign{\smallskip}
            0.5  & 0.054 & 0.030  & 0.024 & 0.018 & 0.016 & 0.014   \\
            0.1  & 0.60 & 0.052 & 0.028 & 0.022  & 0.018 &  0.014 \\
            \noalign{\smallskip}
            \hline
            \noalign{\smallskip}
            0  & $>10$ & 7.9 &  0.20 & 0.064 & 0.036 & 0.018   \\
            \noalign{\smallskip}
            \hline
            \noalign{\smallskip}
         \end{tabular} 
         \tablefoot{{The upper limits of $-\dfeh$} illustrated by the solid lines in Fig\,\ref{fig:Teff-Zsurf-planetacc}.
            $\dMO=0.03\,\Mfin$ and $X\sub{D}=28\,\mathrm{ppm}$.
            $\xi<0.1$ {would} be rare.
            }
        \normalsize
            \end{table}

First let us focus on the $\Zs$ -- $\teff$ trend. As described in Sect.\,\ref{sec:Zsurf}, we find that $\Zs$ of hot stars drops at a certain temperature as a consequence of the dilution (i.e., accretion of metal-free gas due to planet formation). This feature matches the observed trend qualitatively. However, Figs.\,\ref{fig:Teff-Zsurf-planetacc}a--c show that except in the extreme and unrealistic $\xi=0$ case, planet formation cannot lead to a slope that matches the observed one. We conclude that the  planet formation scenario that we are proposing cannot explain the  $\Zs-\teff$ trend seen in the Hyades. 

Let us now examine whether or not planet formation could affect the scatter in $\Zs$ -- $\teff$ and take precedence over another unidentified process. 
As described previously (see Sect.\,\ref{sec:Zsurf}), we expect planet formation to vary from star to star therefore creating a natural scatter of $\Zs$ in a cluster.
Again, for the expected values $\xi\ga0.1$ (see Sect.\,\ref{sec:constraints}), the scatter (which indeed is expected to increase with effective temperature) is much smaller than the observed one (see Table\,\ref{tab:feh}).
In the extreme case of $\xi=0$ (\figref{fig:Teff-Zsurf-planetacc}d), the scatter can be $\sim0.2$\,dex at 6000\,K and even several dex in $\teff\ga6500\,\mathrm{K}$ and therefore most stars are enclosed by the solid line (i.e., the lower limit of $\Zs$ by the dilution) and 0 (i.e., no planet formation).
However, the stars formed with $\xi<0.1$ must be rare (see \citetalias{Kunitomo+17}), and therefore this does not appear to be a viable explanation.

From the considerations above, we conclude that neither the dilution nor the pollution 
{can explain the slope or the scatter} of $\Zs$ of the Hyades cluster.

Recently \citet{Takeda+17} reported that the stars in the 0.1-Gyr-old Pleades cluster have a homogeneous [Fe/H] unlike in the 0.6-Gyr-old Hyades cluster.
Moreover, the iron abundances of stars are homogeneous in the Chamaeleon I star-forming region and the pre-MS cluster $\gamma$ Velorum \citep{Spina+14b,Spina+14a}.
The fact that the metallicity trend is seen only in the older cluster may point to the importance of the element diffusion, which occurs on a long timescale.

\subsection{Implications for chemically inhomogeneous binary systems} \label{sec:binary}

   \begin{table*}[!tb]
      \caption[]{Chemical inhomogeneity in binary systems with planets or debris disks.}
         \label{tab:binary} \small
          \centering
         \begin{tabular}{r|ccccccl}
            \hline
            \hline
            \noalign{\smallskip}
            Name & $\Delta\mathrm{[Fe/H]}$ & Planet & $\teff$ &  $\Mstar$ & Dilution? & Distance & References \\
            \  & [dex] & & [K] & [$\Msun$] & & [au] & \\
            \noalign{\smallskip}
            \hline
            \noalign{\smallskip}
            16 Cyg A & $+0.047$  & -                     &  5830  & {1.05} & consistent & 755 & 1, 2, 3   \\
            B             & 0               & $2.4\,\Mjup$   &  5751  & {1.00}  &              &           &    \\
            \noalign{\smallskip}
            \hline
            \noalign{\smallskip}
            XO-2 N    & $+0.054$ & $0.6\,\Mjup$ & $\approx5300$ & {0.97} & consistent  & 4600 &  4 \\
            S              & 0              & $0.3+1.4\,\Mjup$ & $\approx5300$ & {0.98} &        &  &                 \\
            \noalign{\smallskip}
            \hline
            \noalign{\smallskip}
            $\zeta^{1}$ Ret & $+0.02$ & -                 & 5710 & {0.96} & consistent  & 3713 &  5, 6 \\
            $\zeta^{2}$ Ret & 0            & debris disk & 5854 & {0.99} &               &  &                 \\
            \noalign{\smallskip}
            \hline
            \noalign{\smallskip}
            WASP-94 A & $+0.01$ &  $0.45\,\Mjup$ & $\approx6200$ & { 1.29} & inconclusive  & 2700 &   7, 8  \\
                            B & 0              & $0.62\,\Mjup$ & $\approx6100$ & {1.24} &         &  &                 \\
            \noalign{\smallskip}
            \hline
            \noalign{\smallskip}
            HD133131 A & $-0.03$ & $1.43+0.63\,\Mjup$ & $\approx5800$ & {0.95} & inconclusive  & 360 &  9 \\
                              B & 0              & $2.5\,\Mjup$            & $\approx5800$ & {0.93} &                &  &                 \\
            \noalign{\smallskip}
            \hline
            \noalign{\smallskip}
            HAT-P-4 A & $+0.11$ & $0.7\,\Mjup$ & 6036 & {1.24} & inconsistent  & 28446  &  10 \\
                          B & 0              & -                  & 6037 & {1.17} &                       &             &                 \\
            \noalign{\smallskip}
            \hline
            \noalign{\smallskip}
             HD 80606 & $+0.013$ & $4\,\Mjup$ & $\approx5600$ & {1.03} &  inconsistent  & 1200 &  11 \\
                   80607 & 0              & -                 & $\approx5500$ & {1.00} &         &  &                 \\
            \noalign{\smallskip}
            \hline
            \noalign{\smallskip}
         \end{tabular}
         \tablefoot{$\Delta\mathrm{[Fe/H]}$ is a difference in metallicity and $\Mjup$ is the Jovian mass.
          {\bf References.}
            ${(1)}$ \citet{Tucci-Maia+14};
            ${(2)}$ \citet{Plavalova+Solovaya13};
            ${(3)}$ \citet{Ramirez+11};
            ${(4)}$ \citet{Damasso+15}; 
            ${(5)}$ \citet{Saffe+16};
            ${(6)}$ \citet{Takeda+07}; 
            ${(7)}$ \citet{Teske+16a}; 
            ${(8)}$ \citet{Neveu-VanMalle+14};
            ${(9)}$ \citet{Teske+16b};
            ${(10)}$ \citet{Saffe+17};
            ${(11)}$ \citet{Liu+18}.
                                     }
        \normalsize
            \end{table*}

Binary stars are probably formed from the same molecular cloud core and have the same primordial composition. This is generally the case, but some show differences in metallicity. We focus here on systems in which planets or signs of planet formation have been found to test the validity of our simple planet-formation scenario. 
Table\,\ref{tab:binary} summarizes the observations obtained for seven systems and whether they are consistent or inconsistent with our dilution scenario. 

In three cases, the observed metallicity difference is consistent with the predictions and the idea that efficient (giant) planet formation led to the retention of heavy elements in the planets and a metal-poor star: 
{for} 16 Cyg and XO-2 the star with a planet or more planets is also metal-poor. For $\zeta$ Ret, the metal-poor star is the one with a debris disk. Quantitatively, the differences in [Fe/H] for 16 Cyg and XO-2 are two to three times larger than expected (see Table\,\ref{tab:feh}). This may be explained either by invoking a value of $\xi$ slightly below $0.1$, a value of $X\sub{D}$ closer to 20\,ppm, or by increasing $\dMO$ over our fiducial value $150\,{\Mearth}(\Mstar/\Msun)(Z/\Zsun)$ (see Sect.\,\ref{app:Mrock}). Concerning the latter, we may note that some hot Jupiters appear to be extremely dense and require masses in heavy elements nearing or exceeding $100\,\Mearth$ \citep{Guillot08,Moutou+13,Thorngren+16}. This may be a sign that planet formation was even more efficient than in the solar system. 

However, other cases in Table\,\ref{tab:binary} seem inconsistent with our dilution scenario: HD~80606 is metal-rich but has a known massive eccentric giant planet whereas its binary HD~80607 is metal-poor and has no detected planet. 
HAT-P-4 A hosts a giant planet and is more metal-rich by 0.11\,dex than HAT-P-4 B that does not (both are F stars with an effective temperature above 6000\,K). 

Unfortunately, it is not possible to conclude one way or another at this point for several reasons: 
{the} number of objects for which this measurement is possible is small, we cannot exclude that a massive but far-away planetary system exists but was not detected and planet formation is probably extremely complex and stochastic. We also point out that the determination of differences in metallicity to this level of accuracy is still extremely challenging. For example, \citet{Takeda05} and \citet{Schuler+11} reported no significant metallicity difference in the two components of 16 Cyg, contrary to \citet{Tucci-Maia+14} and \citet{Ramirez+11}.

We also note that \citet{Oh+18} reported that the HD 240430/240429 system has a difference in [Fe/H] of  0.2\,dex  even though no planet has been detected in this system.
They invoked a putative planet which would have been engulfed by the metal-rich star, HD 240430. As we discussed, this seems unlikely, owing to the rapid mixing expected from fingering convection, but indeed we might be observing that particular system while the planet has not yet been completely ingested by its star.

It is therefore difficult {at the moment} to {clearly} attribute differences in {the compositions} of binary stars to {planet formation. This cannot be excluded either}.

\subsection{Implications for $\lambda$ Boo stars} \label{sec:lamboo}
The significant decrease of $\Zs$ of high-mass stars has an important implication for $\lambda$ Boo stars, which make up about 2\% of A-type stars \citep{Gray+Corbally98,Paunzen+01}.
They have lower abundances of Fe-peak (therefore refractory) elements by up to 2\,dex but near-solar abundances of volatile elements.

It is hypothesized that the filtration of dust by planetary objects induces their anomalous compositions \citep[e.g.,][]{Venn+Lambert90,Kama+15}.
Figures\,\ref{fig:M-Zsurf}a and \ref{fig:Teff-Zsurf-planetacc}b show that A-type ($\ga1.4\,\Msun$ and $\ga 7000\,$K) stars formed with a low-entropy ($\xi\la0.1$) accretion can have a significantly refractory-poor surface composition as a consequence of the dilution, supporting this hypothesis.
This matches the observations; namely, the lower limit of $\teff$ of the $\lambda$ Boo stars is also $\sim7000\,\mathrm{K}$ \citep{Murphy+Paunzen17}.

Therefore planet formation around high-mass stars that have been formed with a low-entropy accretion ($\xi\la0.1$, which is rare) could be one potential origin of the peculiar composition of the $\lambda$ Boo stars.
This should be studied more quantitatively in future.
We also note that this is not necessarily a unique solution; namely, there may exist multiple channels producing $\lambda$ Boo stars \citep{Murphy+Paunzen17}.

\section{Planet formation and the solar composition anomaly} \label{sec:discussion-solar}

Finally, in this section, we discuss the solar composition anomaly problem, which has recently been  discussed both theoretically \citep[][]{Chambers10, BC10} and observationally \citep[e.g.,][]{Melendez+09,Ramirez+09,Gonzalez+10,Gonzalez-Hernandez+10,Gonzalez-Hernandez+13, Schuler+11a, Adibekyan+14, Adibekyan+16b, Onehag+14, Nissen15, Spina+16a}.
We revisit this problem focusing on the consequences of planet formation with up-to-date pre-MS evolution models.

\subsection{Background}\label{sec:background-solar}

\citet{Melendez+09} obtained precise surface compositions of the Sun and 11 solar twins, stars almost identical to the Sun.
They found that the solar surface composition is indeed different from most solar twins: the Sun is depleted in refractory elements by up to $\approx0.08\,{\mathrm{dex}}$ (20\%) compared to the average of the solar twins.
\citet{Ramirez+09} confirmed this finding with 22 additional solar twins and found $\sim 15\%$ of them to also have a Sun-like, refractory-poor composition.

\citet{Chambers10} showed that adding $4\,\Mearth$ rocks to the solar surface compensates for the solar composition anomaly, that is, the solar surface composition becomes similar to the average of the solar twins.
Interestingly, the mass of the primordial asteroid belt was $\sim2\,\Mearth$ (see Sect.\,\ref{app:Mrock}) and hence the total mass of the rocky materials in the early inner solar system was $4\,\Mearth$ rocks, which meets the requirement.

However, \citet{Chambers10} used the internal structure of the present-day Sun, whose CZ is much thinner than that of a $1\,\Msun$ pre-MS star (see Sect.\,\ref{sec:int-evols}).
As described in Sect. \ref{sec:intro}, the stellar internal structure is crucial in the dilution scenario.
Since the dilution mechanism should occur only during the pre-MS phase, we revisit their hypothesis on the solar composition anomaly using our evolutionary models of pre-MS $1\,\Msun$ stars.

\subsection{Evaluation of the rock mass required to explain the solar composition anomaly}\label{sec:MCZeff}

Using the internal structure evolutions in Sect.\,\ref{sec:int-evols},
we evaluate the mass of rocks, $\Mrockreq$, required to create the compositional differences between the Sun and solar-twins.
{We do not attempt to match the precise composition of the Sun. 
Instead, we treat H, C, N and O as ``ices'' and refractory material (with higher condensation temperatures) as ``rocks'' following \citet{Lodders03}.
}
\citet{Chambers10} found, {using a MS solar model ($\MCZ=0.025\,\Msun$)}, that the required rock mass is $4\,\Mearth$. {Because we account for stellar evolution, we must scale this value as a function of $\MCZeff (\tp)$, an effective convective zone mass that depends on the time at which heavy elements start being retained in the disk: }
\begin{equation} \label{eq:Mrockreq}
\Mrockreq=4\,{\Mearth} \brafracket{\MCZeff}{0.025\,\Msun}\,.
\end{equation}

\begin{figure}[!t]
  \begin{center}    
        \includegraphics[width=\hsize,keepaspectratio]{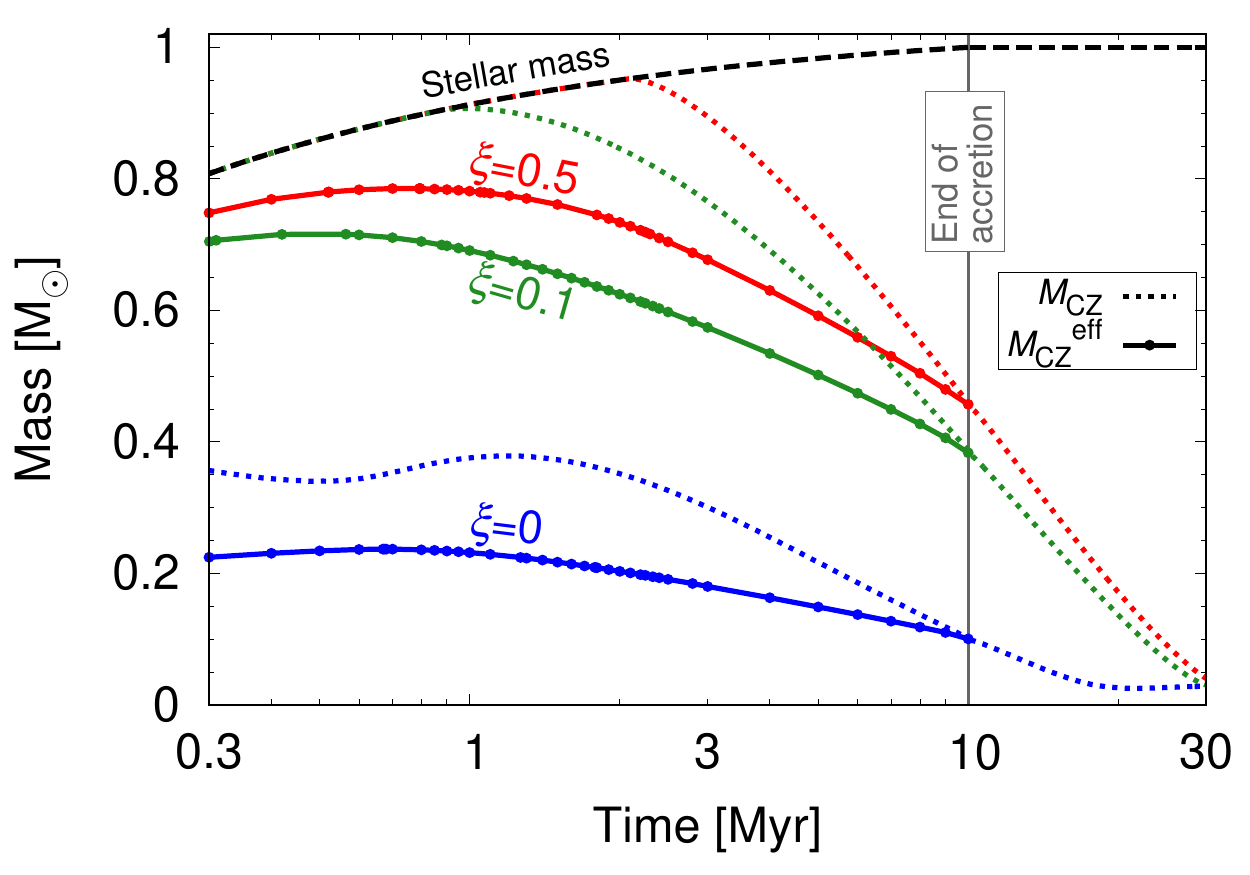}
        \caption{\small{
        Evolution of the convective zone mass ($M\sub{CZ}$, {dotted} lines) and the effective convective zone mass calculated with Eq.\,(\ref{eq:MCZeff}) ($\MCZeff$, solid).
        From top to bottom, $\xi=0.5, 0.1$ and 0.
        The {dashed} line illustrates the evolution of stellar mass and $\Mfin=1\,\Msun$ in all cases.
        }}\label{fig:MCZeff}
    \end{center}
\end{figure}

We derive $\MCZeff (\tp)$ as follows. {We use our simple accretion model (Sect.\,\ref{sec:Za} and Fig.\,\ref{fig:comp-Zs}), but consider that after an initial phase with $\Za(t<\tp)=\Zini$, planet formation suppresses the accretion of heavy elements so that $\Za(t\ge\tp)<\Zini$. We use our approximate solution to calculate $\Zs$ as a function of $\Zini$ and $\tp$. We then derive $\MCZeff(\tp)$ as the CZ mass of a static model for which the accretion of a mass $\dMO$ ($=M\sub{disk}$ at $t=\tp$) with metallicity $\Za(t\ge\tp)$ would give the same surface composition: } 
\begin{equation} \label{eq:MCZeff}
\MCZeff (\tp) =\dMO \frac{\Zs-\Za}{\Zini-\Zs},
\end{equation}
which results from Eq.\,(\ref{eq:simpleZs}).
Figure\,\ref{fig:MCZeff} illustrates the $\MCZeff (\tp)$ derived with $\xi=0.5, 0.1$ and 0 and $\Za(t\geq\tp)=0$. 
As described in Sect.\,\ref{sec:int-evols}, younger stars formed with a larger $\xi$ value have a larger $\MCZeff$.
We note that we also derived $\MCZeff$ with a non-zero $\Za$ but it does not depend on the assumed $\Za$.
{Because $\MCZeff(\tp)$ is an average of $\MCZ(t)$ on the time period from $\tp$ to the end of accretion (10\,Myr), $\MCZeff(10\,{\mathrm{Myr}})=\MCZ(10\,{\mathrm{Myr}})$.}

Using the derived $\MCZeff$, we obtain $\Mrockreq$, which range between 16 and $130\,\Mearth$, depending on $\xi$ and $\tp$ (see Table\,\ref{tab:solutions}).
In Table\,\ref{tab:solutions} the minimum value of $\tp$ is 0.3\,Myr, whereas the maximum values, $t\sub{p,max}$, correspond to the case where after $t\sub{p,max}$ all rocks in the disk are retained in planetesimals/planets and no rocky material accretes onto the star (see discussion in Sect.\,\ref{sec:fir}).
The $\Mrockreq$ value increases with $\xi$.
We note that, with $\xi$ being fixed, the smallest values of $\MCZeff$ and $\Mrockreq$ are obtained for $\tp=t\sub{p,max}$.

We emphasize that, even with the extreme case ($\xi=0$), at least $16\,\Mearth$ rocks are needed to explain the solar anomaly problem.
The $\Mrockreq$ values are much larger than $4\,\Mearth$, that is, the requirement of \citet{Chambers10}.
Moreover, they exceed the possible upper limit of the total rock mass in the early inner solar system, $\sim4\,\Mearth$ (see Sect.\,\ref{app:Mrock}).
Therefore we conclude that the formation of terrestrial planets and asteroid belt alone cannot explain the solar composition anomaly.

\subsection{Ice-to-rock ratio of the solar-system planets}
\label{sec:fir}

Inside of the snow line, we may find at most $4\,\Mearth$ of rocks.
However, beyond that line, much more solids are retained in planets as a mixture of ices and rocks.
As described in Sect.\,\ref{app:Mrock}, we estimate that the total condensates retained in the planetary objects, $\mzp$, is $\sim 150\,\Mearth$, which exceeds the maximum value of $\Mrockreq, 134\,\Mearth$.
Explaining the solar composition anomaly through the accretion of planets is therefore possible but requires that condensates retained in the disk have a rock-rich composition even beyond the snow line. 
In this subsection we pursue this possibility and constrain $\fir$, the bulk ice-to-rock ratio in the solar-system planets required to account for the observed solar composition anomaly, which is testable with observations.

We must therefore consider that after planetesimals and planets start forming ($t\ge\tp$), rocks are retained more efficiently than ices. Within this scenario, the protosun would still be accreting vaporized ices and a significantly reduced amount of rocks. 
Given that $\MCZeff\gg\mzp$, we consider that the primordial ice-to-rock ratio (i.e., that in the primordial molecular cloud core that formed the Sun) is the same as the protosolar ice-to-rock ratio derived from both the evolutionary models of settling and the current measurements of the solar photosphere and meteorites \citep[see][]{Lodders03}\footnote{
The dilution effect implies that the primordial composition of the molecular cloud core differs slightly from the protosolar composition inferred from observations and models \citep{Lodders03}.
  This difference is globally of the order of 5\% (see Fig.\,\ref{fig:comp-Zs}) but could be larger ($\sim 10\%$) for refractory elements. Although not insignificant, this is much smaller than the uncertainties considered in this work.
}.

{Let us derive $\Mrocktot$, the total amount of rocks retained in planets or ejected from the solar system. In order to explain the solar composition anomaly, the total condensate mass, $\mzp$, must consist of $\Mrockreq$ of pure rocks and $\mzp-\Mrockreq$ of a mixture of ices and rocks with a protosolar composition}. 
The latter does not induce an extra compositional anomaly.
According to \citet{Lodders03}, the ice-to-rock ratio of the {solar-system} composition is $f_{\mathrm{ice/rock},\odot}=2.04$ (i.e., 32.9\% are rocks and 67.1\% ices in mass among condensates), and a primordial rock-to-gas mass ratio, $\Zrockini$, is 0.489\%.
We therefore derive
\begin{equation} \label{eq:mrock}
\Mrocktot=\Mrockreq + 0.329 \times(\mzp-\Mrockreq)\,.
\end{equation}

Table\,\ref{tab:solutions} provides the range of values of $\Mrocktot$ compatible with the solar composition anomaly, assuming $\mzp=150\,\Mearth$, for different $\xi$ values and for $\tp$ ranging from 0.3\,Myr to $t\sub{p,max}$.
We recall that for $\tp=t\sub{p,max}$
all rocks in the disk ($\mdr \equiv M\sub{disk}\,\Zrockini$ in mass) are completely retained in planetesimals and planets and no rocky material is accreted by the central star (after $\tp$).
For $\tp>t\sub{p,max}$, we need $\Mrocktot>\mdr$, which is impossible.
Conversely, lower $\tp$ values ($< t\sub{p,max}$) are plausible. They correspond to a partial retention of rocks into planets when $t > \tp$.
In this case, contrary to the approach in Sect.\,\ref{sec:discussion}, we do not hold $\dMO$ fixed but allow it to vary.

The values of $t\sub{p,max}$ decrease for increasing $\xi$. As described in Sect.\,\ref{sec:MCZeff}, stars formed with a larger $\xi$ have a thicker CZ, which requires the accretion of a larger amount of rock (metal)-free gas to be diluted. Therefore, more rocks should be retained in planets or planetesimals, namely, a larger $\Mrocktot$ is required. For this, since the disk rock mass $\mdr$ decreases with time, planet formation starts at progressively younger ages.
For values of $\xi$ allowed by \citetalias{Kunitomo+17}, we obtain that $t\sub{p,max}=1.10-1.22$\,Myr, that is, efficient retention of rocks by the planet formation process must begin no later than 1.22\,Myr after the beginning of the formation of the protosun.

The smallest values of $\Mrocktot$ are obtained for $\tp=t\sub{p,max}$.
We thus obtain that the amount of rocks in planets for $\xi=0.1$--0.5 is $\Mrocktot=90$--$134\,\Mearth$ ($60$--$75\,\Mearth$ when considering $\xi=0$). The remainder (i.e., $\mzp-\Mrocktot$) is made of ices ($M\sub{ice}=16$--$60\,\Mearth$ for $\xi=0.1$--0.5 -- or up to $90\,\Mearth$ for $\xi=0$).
From these, we obtain the bulk ice-to-rock ratios in {both the solar-system planets and ejected objects}, $\fir\equiv M\sub{ice}/M\sub{rock}$, to explain the solar compositional anomaly. {(We assume that the ejected planets and planetesimals have on average the same composition as the remaining ones.)}

   \begin{table}[!tb]
      \caption[]{Solutions required to explain the solar composition anomaly.}
         \label{tab:solutions} \small
         \begin{tabular}{p{0.4cm}|p{1.3cm}p{1.2cm}p{1.1cm}p{0.9cm}|p{1.3cm}}
            \hline
            \hline
            \noalign{\smallskip}
            $\xi$ & $\MCZeff$ & $\Mrockreq$ & $\Mrocktot$ & $t\sub{p,max}$ & $\fir$ \\
             & $[\Msun]$ & $[\Mearth]$ & $[\Mearth]$ & [Myr] & \\
            \noalign{\smallskip}
            \hline
            \noalign{\smallskip}
            0.5  & 0.46--0.79 & 74--130 & 99--134 & 1.10 & 0.12--0.16  \\
            0.1  & 0.38--0.72 & 61--115 & 90--127 & 1.22 &  0.19--0.23 \\
            \noalign{\smallskip}
            \hline
            \noalign{\smallskip}
            0  & 0.10--0.24 & 16--38 &  60--75 & 2.33 & 1.01--1.13 \\
            \noalign{\smallskip}
            \hline
            \noalign{\smallskip}
         \end{tabular} 
         \tablefoot{We adopt $\mzp = 150\,\Mearth$.}
        \normalsize
            \end{table}

\begin{figure}[!t]
  \begin{center}    
        \includegraphics[width=\hsize,keepaspectratio]{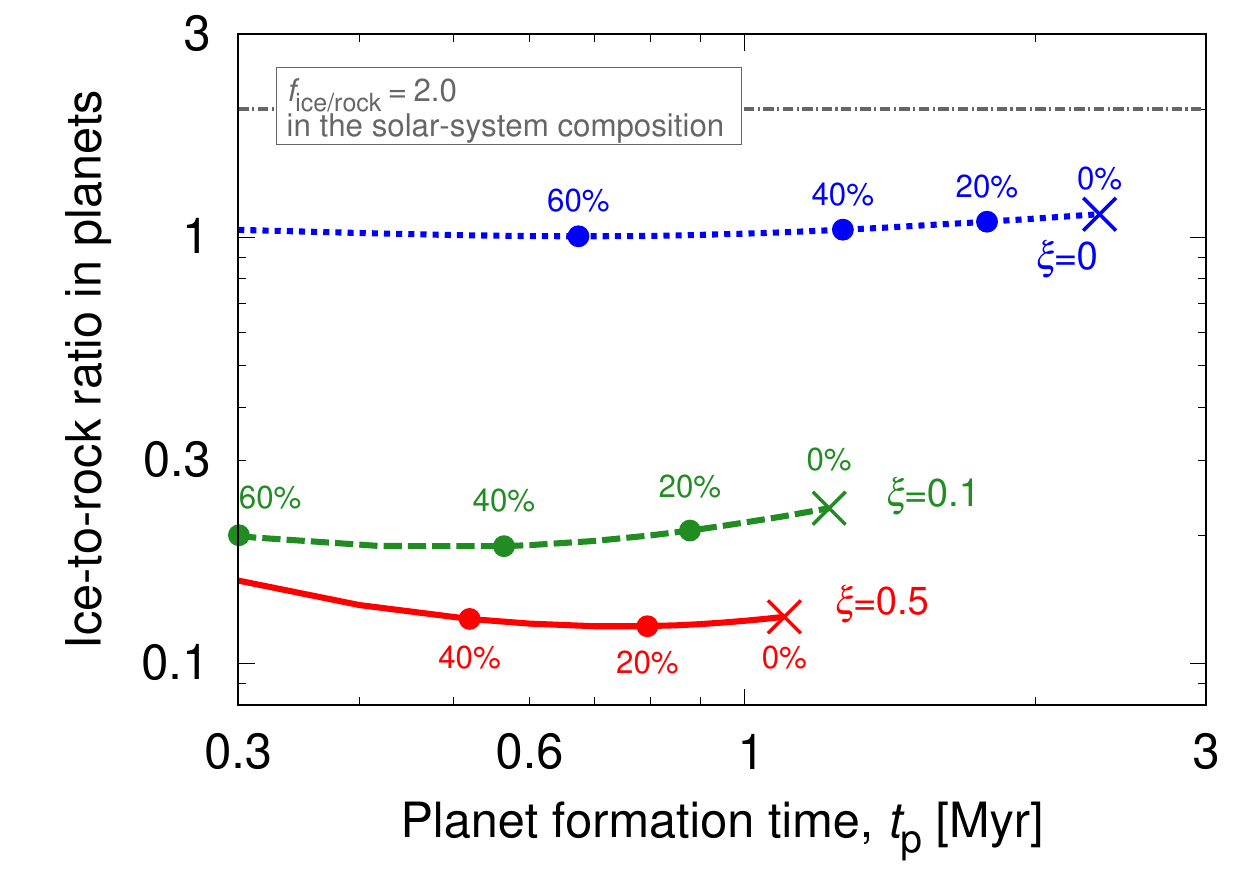}
        \caption{\small{
        Ice-to-rock ratios in solar-system planets required to explain the solar compositional anomaly {as a function of $\tp$, the time at which efficient planet formation starts and suppresses the accretion of rocks (metals) onto the star. The lines correspond to different values of the heat accretion parameter $\xi=0.5, 0.1$ and $0$, from top to bottom. Percentage figures (0 to 60\%) indicate the fraction of rocks still in the accreting flow of the protosun for $t\ge\tp$ (100\% corresponds to a rock-mass fraction equal to the protosolar value, $0.00489$). The calculations assume $\mzp=150\,\Mearth$.}
        }}
        \label{fig:Mrock}
    \end{center}
\end{figure}

Figure\,\ref{fig:Mrock} shows how $\fir$ varies as a function of $\tp$. The crosses illustrate the solutions for $\tp=t\sub{p,max}$.
The rock-to-gas ratio of the accreted matter is 
\begin{equation}
\Zrockacc=\Zrockini-{\Mrocktot}/{\dMO}\,.
\end{equation}
$\Zrockacc/\Zrockini$ therefore corresponds to the fraction of rocks still present in the accretion flow. 
Symbols in Fig.\,\ref{fig:Mrock} pinpoint the fraction, $\Zrockacc/\Zrockini$, ranging from 0 to 60\% ($\xi\ge 0.1$).

Figure\,\ref{fig:Mrock} shows that $\fir\approx0.1, 0.2$ and 1.1 for $\xi=$ 0.5, 0.1 and 0, respectively.
The solutions are almost insensitive to $t\sub{p}$.
These values are lower than 
$f_{\mathrm{ice/rock},\odot}=2.04$. This means that somehow planets should retain rocks efficiently.

Let us examine {how different values of $\mzp$ affect the solutions}.
We consider two cases with $\mzp=97$ and $168\,\Mearth$ as a lower and upper limit in our solar system (Sect.\,\ref{app:Mrock}).
For $\mzp=168\,\Mearth$, the solutions are $M\sub{rock}=139, 128, 76.7\,\Mearth$ and $\fir=0.21, 0.39, 1.19$ for $\xi=0.5, 0.1$ and 0, respectively.
With $\mzp=97\,\Mearth$, $M\sub{rock}=50.9\,\Mearth$ and $\fir=0.91$ for $\xi=0$ and there is no solution for $\xi \geq 0.1$.
Therefore, together with our results in \citetalias{Kunitomo+17}, there is no realistic solution in the latter case.
In all cases, explaining the solar composition anomaly with planet formation requires that the solar-system planets have globally an ice-to-rock ratio that is significantly below the solar value.

\subsection{Discussion} \label{sec:discussion-discussion-solar}

The solar ice-to-rock ratio, 2.04, is much larger than the values that we obtain to explain the solar composition anomaly with planet formation (e.g., 0.23 if  $\mzp=150\,\Mearth$ and $\xi=0.1$).
However, we stress that the global ice-to-rock ratio in planets of the solar system has been inferred from the solar value, but not obtained independently.
Its value in the giant planets, which are by far the largest contributors, is unknown \citep[e.g.,][]{Guillot05}.
Furthermore, several objects in the solar system have values much lower than the solar one: 
{for} example it has been inferred to be 0.64 in Enceladus \citep{Porco+06}
(about 61\% of the mass of Enceladus is rocks)
and 0.53 in Pluto \citep{McKinnon+17}.
In comet 67P/Churyumov-Gerasimenko, in situ measurements found a gas-to-dust ratio (a proxy for the ice-to-rock ratio) of 0.17 to 0.5 \citep{Rotundi+15}.
It is possible that the giant planets were formed mostly from ice-poor planetesimals inside of the snow line and that the overall ice-to-rock ratio in planets is therefore much smaller than that in the Sun \citep{Ida+Guillot16}.
Of course, alternative explanations of the origin of the solar composition anomaly, such as a migration of the Sun in the Galaxy, exist
\citep[see][and references therein]{Adibekyan+17}.

\section{Conclusions}\label{sec:conclusion}

In this article, we explored the consequences of planet formation on stellar surface compositions.
{We examined two opposite effects: the dilution effect due to the low-metallicity of the gas accreted in the last stages of stellar accretion, and the pollution from planetesimals and planets possibly accreted by the central star}.  
In order to quantify these effects, the key ingredients are the thickness of the stellar surface convective zone (CZ) and its evolution in time, the amount of heavy elements retained by the planet formation process, {and the amount and time of the accretion of planetesimals and planets.}

First, following \citetalias{Kunitomo+17}, we analyzed the evolution of stellar interiors on the pre-MS, as controlled by three parameters: 
the efficiency with which heat is injected from the accreting flow to the star, $\xi$, 
the mass fraction of deuterium, $X\sub{D}$, and the stellar mass, $\Mfin$.
The evolution of the CZ can be subdivided in four phases:
\begin{description}
\item[(1)] Initially the convective-radiative structure depends on how accretion energy is distributed within the star and can be highly variable.
\item[(2)] Efficient deuterium burning starts and ensures that the interior becomes largely convective.
\item[(3)] The CZ retreats and a central radiative zone grows essentially due to the progressive increase in temperature and subsequent decrease in opacity of the interior.
\item[(4)] The star reaches the MS, at which point the properties of the CZ become almost time-independent.
\end{description}
These four phases are reached at different ages. 
For likely values of the heat injection parameter ($\xi\wig{>}0.1$ -- see \citetalias{Kunitomo+17}), we found that Phase\,1 lasts $\sim 0.1\,$Myr, that is, when the star is less than half of its final mass.
Phase\,2 lasts for about $1\,$Myr for solar-mass stars, but $10\,$Myr for a $0.5\,\Msun$ star, and is inexistent for a $1.5\,\Msun$ star. 
Similarly, the end of Phase\,3 depends on stellar mass, from about $100\,$Myr for a $0.5\,\Msun$ star, $30\,$Myr for a $1.0\,\Msun$ star, and $10\,$Myr for a $1.5\,\Msun$ star. 
For $\xi\wig{>}0.1$ we found that after Phase\,1, the CZ evolution is relatively close to that of classical (non-accreting) evolution tracks, with a slightly faster evolution towards the MS for lower values of $\xi$ and of $X\sub{D}$.

By causing changes in the composition of accreted material, planet formation will affect stellar composition but the relatively slow retreat of the stellar CZ during the first million years means that the effect will generally be small. 
It should be maximal for high-mass stars (smaller CZ) and {for stars hosting giant planets (requiring large amounts of solids from the early stages of protoplanetary disk evolution).
}

Next, we estimated that in the solar system, between $97$ and $168\,\Mearth$ of condensates formed planets or were ejected from the system. We hence adopted as a fiducial scenario the accretion of $0.03\,\Mfin$ of zero-metallicity gas in the last stages of the evolution of the circumstellar disk (dilution scenario).
This is expected to be an uncertain but reasonable limit for a star with several giant planets like our Sun.
Conversely, stars without planets would not show a significant metallicity change.

Another mechanism, that is, planet or planetesimal engulfment during the MS phase, can increase stellar metallicity. However, this is expected to be quickly erased by thermohaline convection on a 10- to 100-Myr timescale \citep{Theado+Vauclair12}.
We estimate that a significant {($\sim0.1\,\mathrm{dex}$)} increase in metallicity due to pollution by planet and planetesimal engulfment should be limited to massive stars with effective temperatures above $\sim6500\,\mathrm{K}$.

With this fiducial setting (i.e., $\dMO= 0.03\,\Mfin$, $\xi=0.1$ and $X\sub{D}=28\,\mathrm{ppm}$), we found that the dilution scenario results in a decrease of [Fe/H] by up to 0.6\,dex for $\sim7000\,\mathrm{K}$ stars dropping to less than 0.05\,dex at 6500\,K and less than 0.02\,dex at 5500\,K. Given the variability of planet formation and the relative scarcity of systems with giant planets, we estimate that this is representative of the scatter in stellar composition that should be imprinted by planet formation.
This scatter is smaller and has a different $\teff$ dependence than that measured in the Hyades cluster \citep{Takeda+13}, 
indicating either that systematic effects bias the abundance retrievals, 
that observational uncertainties are still too large, 
or that other processes such as stellar diffusion play a larger role.

We examined how dilution can be used to interpret the differences in composition observed in some binary systems with planets or signs of planet formation: 
{several} systems with planets have stellar components differing in metallicity by 0.01 to 0.06 dex, values which are within a factor of approximately two of those estimated on the basis of planets in the solar system (i.e., $\dMO= 0.03\,\Mfin$; see Fig.\,\ref{fig:Teff-Zsurf-planetacc} and Table\,\ref{tab:feh}). Given the ambiguities of the findings, the small number of cases and the complexity and variability of planet formation itself, it is however too early to draw any conclusion at this point, except that it is a promising area of research. 

Other important objects are $\lambda$ Boo stars, 
a small fraction ($\sim2\%$) of A-type stars 
{with effective temperatures above 7000\,K}
with a significant depletion of up to 2\,dex in refractory elements but solar compositions in volatile elements.
{For such temperatures,} the CZ has shrunk rapidly enough to see very large changes in stellar composition due to planet formation, assuming relatively low values of $\xi \sim 0.1$ (see Fig.~\ref{fig:Teff-Zsurf-planetacc} and Table~\ref{tab:feh}). This however requires another ingredient, namely that planet formation preferentially favors the retention of refractory elements over volatile ones. Indeed, observations of accreting Herbig stars indicate that in systems which we expect to host giant planets, the accreting gas is depleted in refractory elements by 0.5\,dex, while the volatile elements show normal abundances \citep{Kama+15}.
Planetesimal formation inside of the ice line can lead to efficient retention of refractory elements and the loss of ices \citep{Ida+Guillot16}.

Finally, we looked for an explanation of the fact that our Sun is poor in refractory elements compared to the average solar twin \citep{Melendez+09, Ramirez+09, Spina+16a}.
The traditional explanation is that this deficit is created by the formation of the terrestrial planets and the proto-asteroid belt \citep{Chambers10}.
We showed however that this explanation is incompatible with the pre-MS evolution of the Sun and the fact that, early on, its surface CZ was much larger than it is now.
Instead, the solar composition anomaly may be linked to the dilution mechanism proposed here.
The early formation of a proto-Jupiter is known to have prevented the Sun from accreting material from the outer disk. If this material had a lower-than-solar ice to rock ratio, this would have led to a depletion of refractory elements in the gas accreted by the Sun in its last stage of formation. 
In order for this scenario to work, we found that the global ice-to-rock ratio of planets should be less than $\approx0.4$ and efficient planetesimal formation should have begun at most about 1\,Myr after the beginning of the formation of a proto-Sun (i.e., a second Larson's core).
These values are much smaller than the protosolar ice-to-rock ratio of $2.04$ \citep{Lodders03} implying that planets in the solar system should be extremely rock-rich. This is not impossible \citep[see][]{Ida+Guillot16} especially given that the giant planets, whose ice-to-rock ratio is unknown, contain most of the planetary mass of heavy elements. 
The scenario is similar, at least qualitatively, to the one needed to explain the composition of $\lambda$ Boo stars.

Planet formation thus affects stellar compositions {in a subtle and complex way. We predict that, in general, the dominant effect should be a slight decrease of the metallicity of stellar outer convective zones compared to that of their deeper interior and to the primordial value. Observations of} compositional differences in some binary stars, the composition anomalies of $\lambda$ Boo stars, and perhaps of our own Sun {do point in that direction, with multiple exceptions
 however}. 
In order to make progress, we therefore need very accurate measurements of the compositions of stars in clusters and those of binary stars and limits on the presence of planetary companions around them. 
We have shown that the solar composition anomaly can be explained only if planets in the solar system are extremely ice-poor. 
Testing this hypothesis requires much better constraints on the composition of giant planets, something Juno has started doing for Jupiter \citep[see][]{Helled+Lunine14, Bolton+17b,Guillot+18} but is eagerly awaited for the other planets, in particular Uranus and Neptune.

\begin{acknowledgements}
We express our gratitude to Andr\'e Izidro for providing simulation results and to John Chambers for providing data. 
We appreciate the constructive comments of the anonymous referee, which helped us to improve this paper.
We are also grateful to Shu-ichiro Inutsuka, Hiroshi Kobayashi, Shinsuke Takasao, Yoichi Takeda, Vardan Adibekyan, and Corinne Charbonnel for fruitful discussions and comments.
M.K. thanks for the hospitality he experienced during his visit to the Observatoire de la C\^ote d'Azur, which was made possible thanks to support from \textit{OCA BQR}.
This work was supported by Grant-in-Aid for JSPS Fellows 24$\cdot$9296, JSPS KAKENHI Grant Numbers 23244027, 15H02065, 16H02160 and 17H01153, JSPS Core-to-Core Program ``International Network of Planetary Sciences'', and the Astrobiology Center Program of National Institutes of Natural Sciences (NINS) (Grant Number AB301023).
\end{acknowledgements}

\bibliographystyle{aa}


\begin{appendix}

\section{Underlying physics of internal structure evolution} \label{sec:physics}

In this appendix, we describe the underlying physics of the pre-MS evolution of stellar internal structure, as governed by their entropy profile, opacity, and mixing of freshly accreted deuterium. Our goal is to explain the evolution of the convective and radiative zones shown in Fig.\,\ref{fig:conv-multi}.

For a homogeneous composition, the convective stability criterion {(i.e., Schwarzschild criterion, $\nabla\sub{rad} < \nabla\sub{ad}$) is} expressed {as a condition on the Rosseland opacity, $\kappa\sub{R}$} \citep[see, e.g.,][]{Kippenhahn+Weigert90}:
\begin{equation}
\kappa\sub{R} < \frac{64\pi}{3} \frac{GM_{r}}{P}  \frac{\sigma\sub{SB} T^{4}}{L} {\nabla\sub{ad}}\equiv\kappa\sub{crit}\,, \label{eq:conv_sch2}
\end{equation}
where $L$ is the luminosity, $T$ the temperature, $P$ the pressure, $\sigma\sub{SB}$ the Stefan-Boltzmann constant, $G$ the gravitational constant, $M_{r}$ the mass coordinate, and $\nabla\sub{rad}$ and $\nabla\sub{ad}$ the radiative and adiabatic temperature gradient ($\nabla=\dd \ln T/\dd \ln P$), respectively. We note that $\nabla\sub{ad}=0.4$ for the monoatomic perfect gas.

\begin{figure}[!t]
  \begin{center}
    \includegraphics[width=\hsize,keepaspectratio]{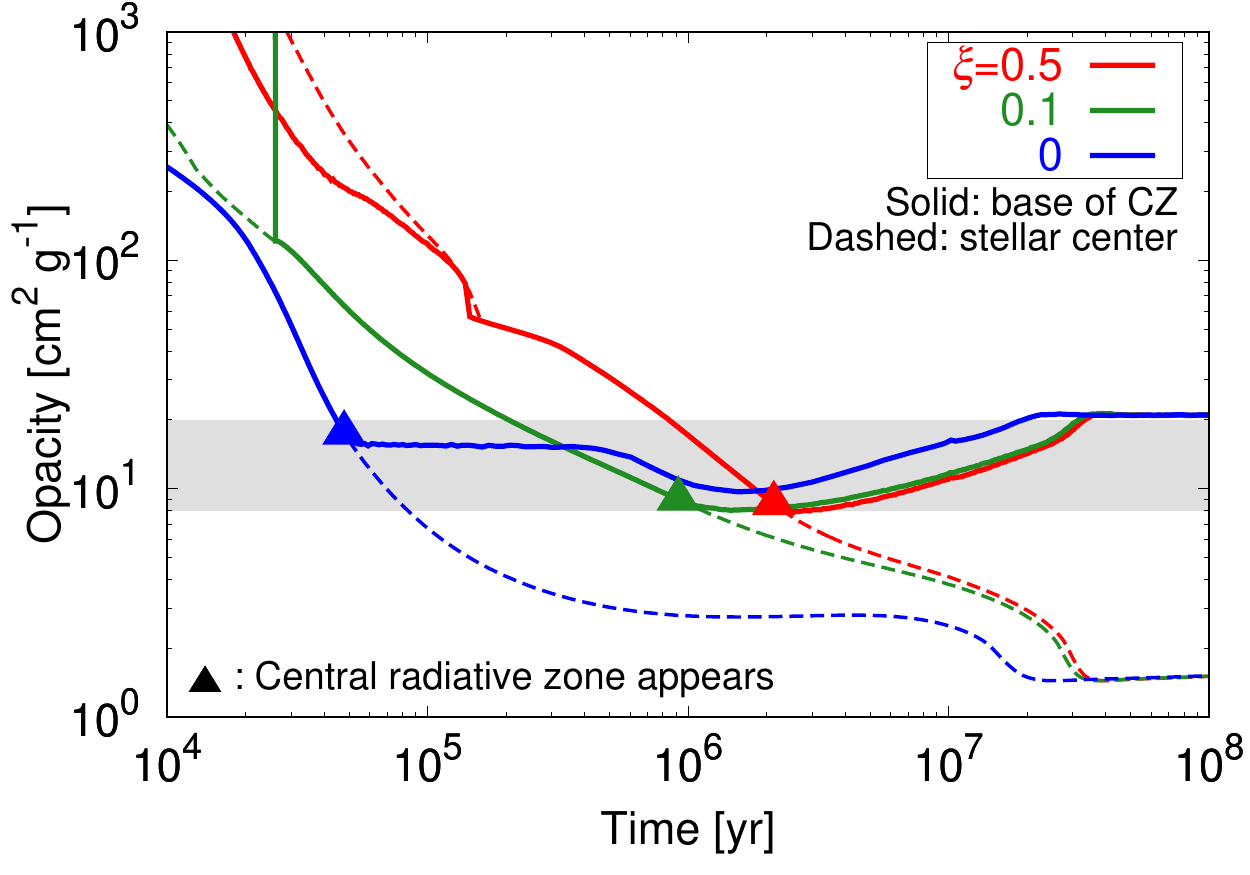}
        \caption{\small{
Temporal evolution of the opacity at the base of a surface CZ (solid lines) and at the center (dotted).
$\xi=0.5$ (red), 0.1 (green) and 0 (blue) from top to bottom 
{and $\Mfin = 1\,\Msun$.}
}}\label{fig:t-kap}
    \end{center}
\end{figure}

Figure\,\ref{fig:t-kap} shows that this critical opacity, $\kappa\sub{crit}$, remains relatively constant around $\approx10\,\mathrm{cm^2\,g^{-1}}$ in time and for different $\xi$ values (i.e., the points in Fig.\,\ref{fig:t-kap} have the almost same value).
The evolution of the opacity in the stellar interior thus plays a crucial role in the growth of the radiative zone. The radiative zone appears when the opacity at the center falls below that critical value. This occurs when the central temperature reaches $\approx2$--$4\,\times10^{6}\,\mathrm{K}$ \citep{Chabrier+Baraffe97}.

Convective stability is also governed by deuterium burning during the protostellar phase because it maintains a high luminosity throughout the interior. 
However, since pre-existing deuterium is quickly depleted, the supply of fresh deuterium by the accreting flow is most important.
This fresh deuterium is delivered down into the stellar interior by convective turbulent mixing.
After a radiative core emerges, the fresh deuterium is no longer delivered to the star's central region but is burned at the base of the CZ \citep[forming the so-called radiative barrier, see][]{Palla+Stahler93}.

\begin{figure}[!t]
  \begin{center}
        \includegraphics[width=\hsize,keepaspectratio]{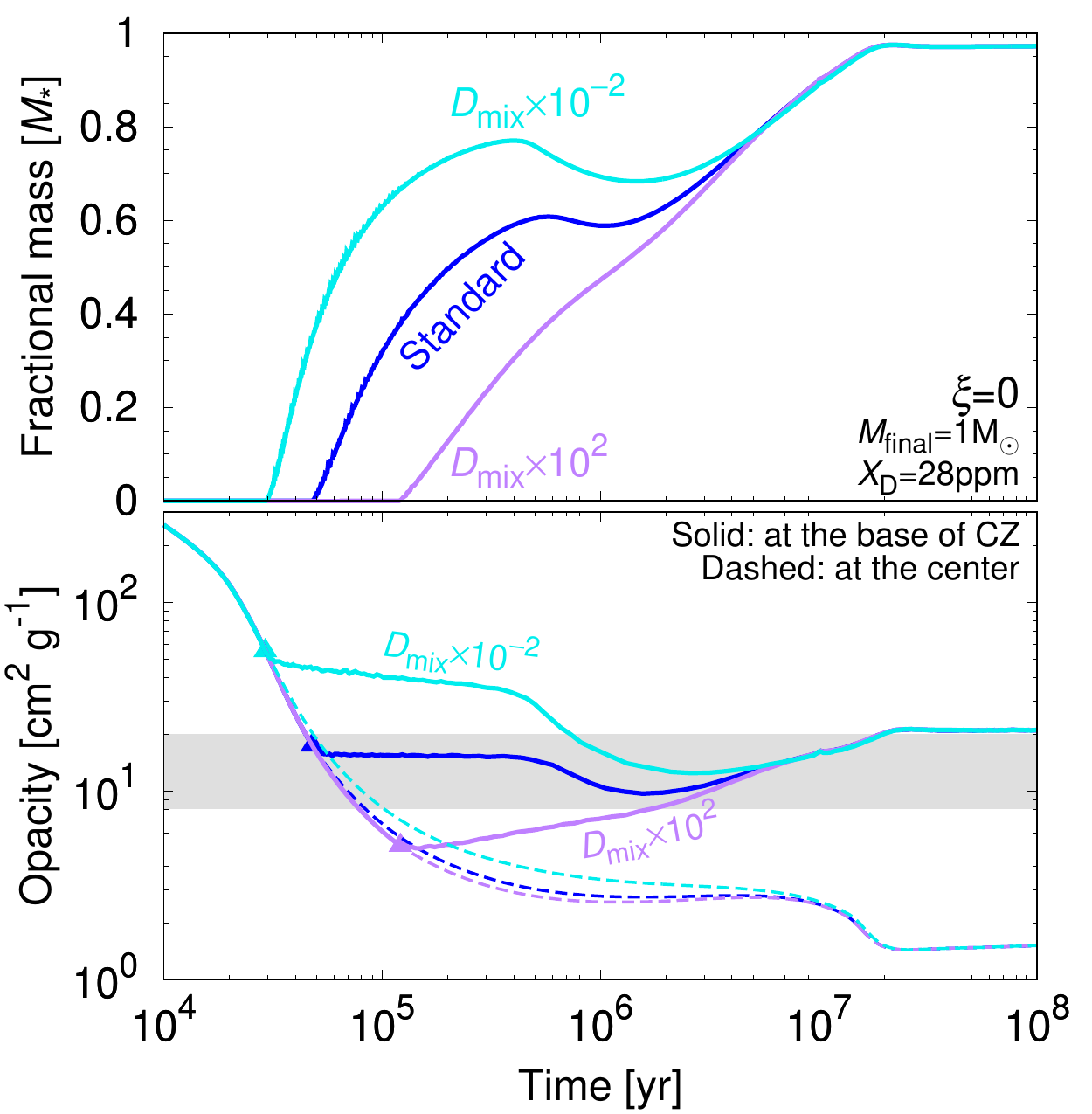}
        \caption{\small{
        Evolution with time of the fractional mass of the central radiative zone (\textit{top panel}) and of the internal opacities (\textit{bottom}), for a $1\,\Msun$ star, assuming a heat injection parameter $\xi=0$ and deuterium mass fraction $X\sub{D}=28\,\mathrm{ppm}$.
        The solid and dashed lines indicate values at the base of the surface CZ and at the center, respectively.
        Three cases are considered: 
        {a} standard case (blue line), one in which the diffusion coefficient of convective mixing is artificially increased (purple) or decreased (cyan) by a factor of 100. 
        The gray area in the bottom panel indicates the range of values of the critical opacity at the base of the CZ in the standard case ($\kappa\sub{crit}\sim10\,\mathrm{cm^{2}g^{-1}}$). The triangles indicate the time at which the radiative zone appears at the center and starts growing. 
}}\label{fig:t-kap-diff}
    \end{center}
\end{figure}

{The importance of deuterium burning is demonstrated in Fig.\,\ref{fig:t-kap-diff}, which shows pre-MS evolutions obtained with a diffusion coefficient of convective mixing, $D\sub{mix}$.
Here we artificially increase or decrease $D\sub{mix}$ over that predicted from mixing length theory by a factor of 100.
Figure\,\ref{fig:t-kap-diff} shows that the} value $\kappa\sub{crit}\sim10\,\mathrm{cm^{2}/g}$ at which the central radiative zone appears is not universal but can be affected by {mixing processes} in protostars.
When diffusion in the convective zone is slower, 
less deuterium is brought to the central region, 
the central luminosity is smaller,
$\kappa\sub{crit}$ is larger,
and therefore the radiative zone appears more rapidly. 
The opposite is true for a larger diffusion coefficient.
The results clearly indicate that the development of a radiative core is controlled by not only the opacity but also the deuterium burning.

{Let us now turn to \figref{fig:conv-multi}. We explain three noteworthy features seen in the different panels: 
(i) a rapid development of a radiative core {for low $\xi$ values}, (ii) an expansion of the surface CZ in the $\xi=0$ case at around $10^{6}\,\mathrm{yr}$, and (iii) in the $m\sub{ke}=0.1$ case, a central region which is always radiative.
}

First and most importantly, \figref{fig:conv-multi} illustrates that the internal structure evolutions strongly depend on $\xi$: 
{stars} with a lower $\xi$ value develop a radiative core more rapidly.
This can be understood as follows.
Such stars have a smaller radius (see \citetalias{Kunitomo+17}).
From the virial theorem, this corresponds to a higher internal temperature.
This results in a lower opacity (cf. Kramers' law; $\kappa\propto T^{-3.5}$).
Therefore stars formed with lower $\xi$ values develop a radiative core more rapidly.

Let us explain (ii), namely the temporary expansion of the surface CZ in the $\xi=0$ case at around 1\,Myr.
Since the stellar interior becomes hot and less opaque at this moment, a large amount of entropy is carried toward the surface region \citep[so-called luminosity wave, ][]{Stahler+86}.
However the surface region is still opaque.
Therefore the entropy accumulates at the base of CZ and then the surface CZ expands for a while (see also Fig.\,B.2 of \citetalias{Kunitomo+17}).

Finally, let us explain {why the non-uniform heat injection model with $m\sub{ke}=0.1$ is characterized with a center that is always radiative (see \figref{fig:conv-multi}d).
This case corresponds to an accretion heat that is deposited entirely in an outer, narrow region. The specific entropy gradient is therefore locally positive during the protostellar phase}. Consequently, the temperature maximum is located slightly off-center and therefore deuterium burning also starts away from the center. As a result, the central region always remains radiative.
This means that the primordial composition {may be} preserved in the central region {and may differ from that in the rest of the star}. This may be testable by asteroseismic observations.

\section{Varying accretion history and initial metallicity} \label{sec:tdisk-zini}

\begin{figure}[!t]
  \begin{center}    
        \includegraphics[width=\hsize,keepaspectratio]{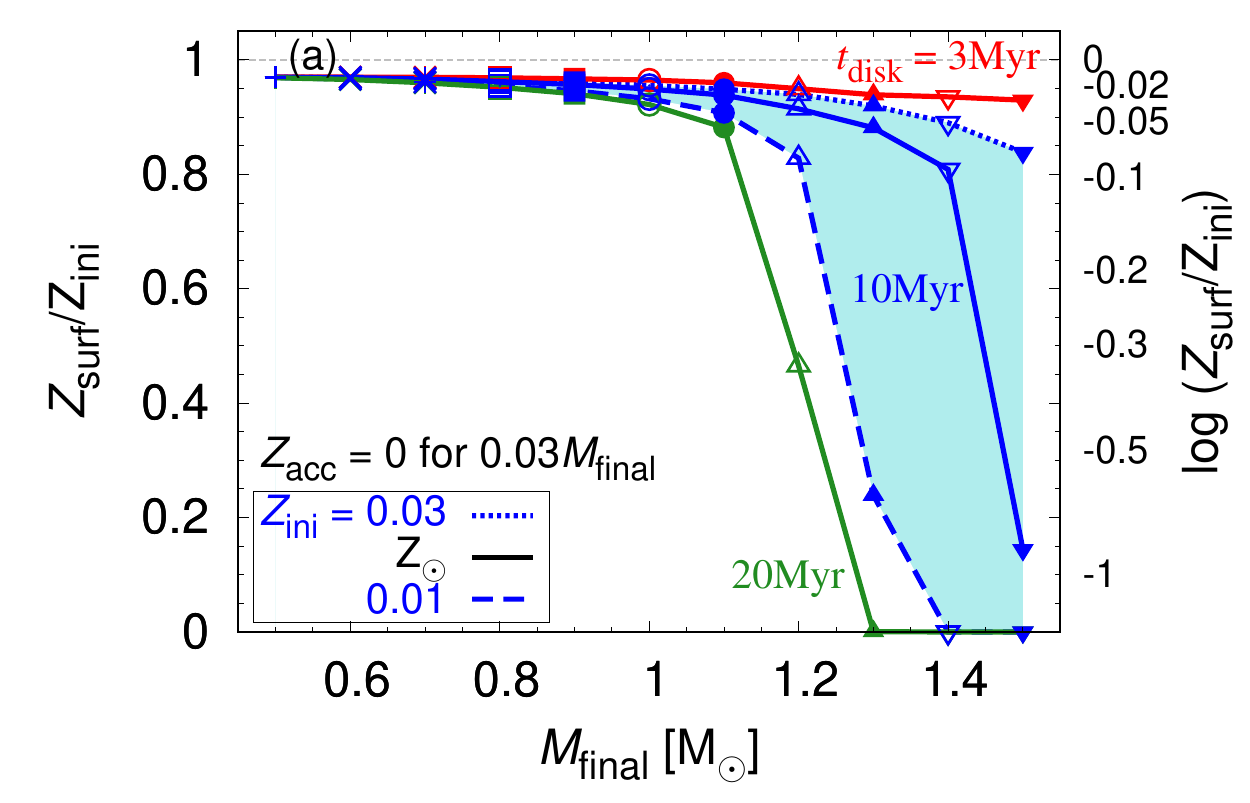}
        \includegraphics[width=\hsize,keepaspectratio]{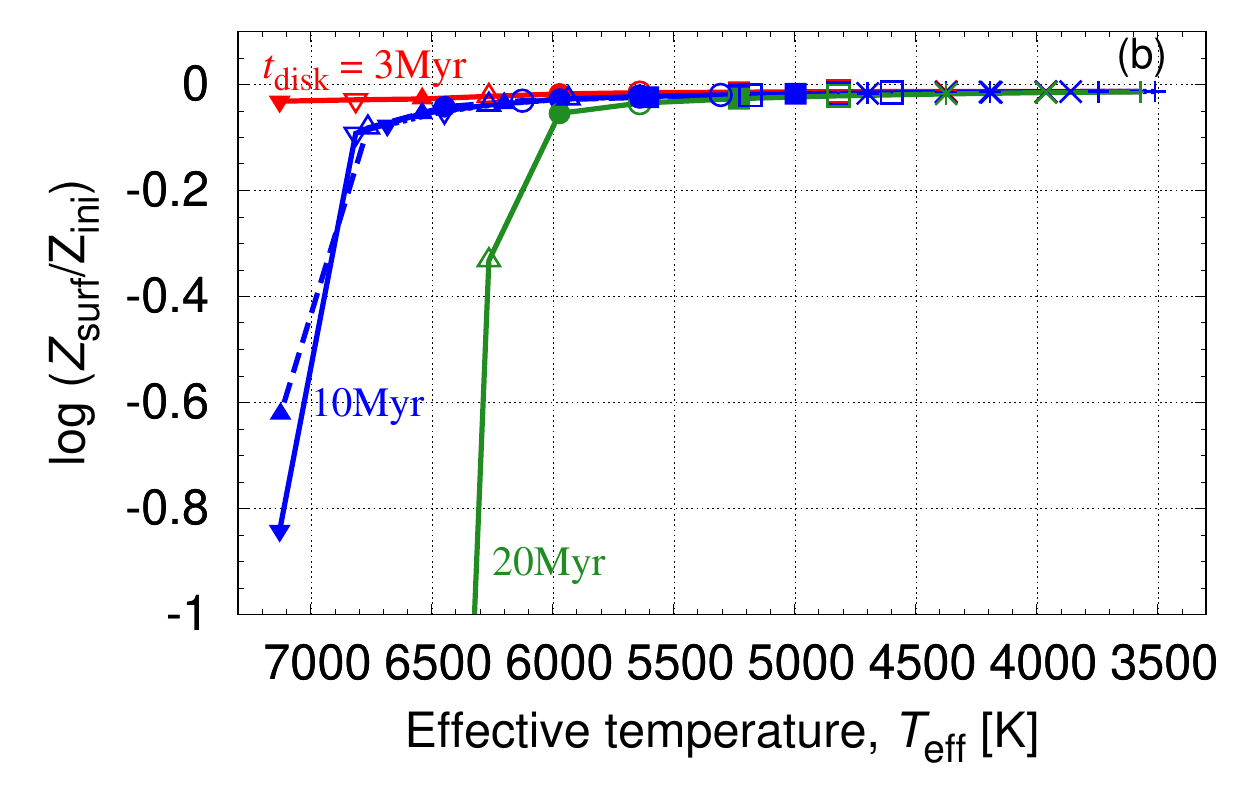}
        \caption{\small{
        The same figures as Figs.\,\ref{fig:M-Zsurf}a and \ref{fig:M-Zsurf}b, but with varying accretion history ($t\sub{disk}$, see text) and initial metallicity, $\Zini$.
        The solid lines indicate the cases with $\Zini=\Zsun (=0.0197)$ and $t\sub{disk}=3$ (red line), 10 (blue) and 20\,Myr (green) from top to bottom.
        The dotted and dashed lines represent the cases with $\Zini=0.03$ and 0.01, respectively, and with $t\sub{disk}=10$\,Myr.
        The other settings are fiducial ones, namely $\dMO= 0.03\,\Mfin$, $\xi=0.1$, the uniform heat distribution and $X\sub{D}=28\,\mathrm{ppm}$).
        In the bottom panel, the cases with $\Zini=0.01$ (blue dashed) and $\Mfin\geq1.4\,\Msun$ are not plotted because $\Zs=0$.
        We adopt $\teff$ at 0.3\,Gyr when all stars are in their MS.
        }}\label{fig:tdisk-zini}
    \end{center}
\end{figure}

Here we discuss the effect of varying accretion history and initial metallicity.
In the main text, we adopted fiducial values for the disk lifetime, $t\sub{disk}=10\,$Myr and for the initial metallicity $\Zini=0.02$.
However there can be star-to-star scatters or systematic offsets in these values. For example the mean metallicity of the Hyades cluster is larger than the solar \citep{Takeda+13}.
Figure\,\ref{fig:tdisk-zini} shows the values of $\Zs$ with varying $t\sub{disk}$ and  $\Zini$: $t\sub{disk}=3$ and 20\,Myr \citep[e.g.,][]{Haisch+01} and $\Zini=0.01$ and 0.03.
The other settings are set to be fiducial ones, namely $\dMO= 0.03\,\Mfin$, $\xi=0.1$, a uniform heat distribution and $X\sub{D}=28\,\mathrm{ppm}$.

Figure\,\ref{fig:tdisk-zini}a shows that metal-poor high-mass ($\ga1.1\,\Msun$) stars are more significantly affected by dilution. 
As we described in Appendix\,\ref{sec:physics}, the shrinkage of CZ is controlled by the opacity (and deuterium burning).
Therefore since higher-mass stars (see also Fig.\,\ref{fig:Mfin-conv}) and metal-poor stars have a lower opacity, they have a thin CZ.

By contrast, in the $\teff$ -- $\Zs$ plane (Fig.\,\ref{fig:tdisk-zini}b), the resultant $\Zs$ values are not sensitive to $\Zini$ values.
This is because both the CZ extension and $\teff$ are sensitive to the metallicity.
Indeed, at a fixed mass, a larger metallicity corresponds to a lower $\teff$ and to a thicker CZ due to a larger opacity.
Therefore, to obtain the same $\teff$, given a larger metallicity, a larger mass is required, and vice versa. This results in a thicker CZ.
The two combined effects almost counterbalance one another, leading to a weak dependency of $\MCZ$ (and therefore $\Zs$) on the metallicity at fixed $\teff$.

As for the variation in $t\sub{disk}$, Figs.\,\ref{fig:tdisk-zini}a and \ref{fig:tdisk-zini}b show that it has a significant impact on resultant $\Zs$ values due to the dilution.
This is because $\MCZ$ in the late phase (0.97--1\,$\Mfin$) are different: it is large in the cases with $t\sub{disk}=3\,\mathrm{Myr}$ ($t\approx1.6$--3\,Myr), whereas small with $t\sub{disk}=20\,\mathrm{Myr}$ ($t\approx4.7$--20\,Myr).
This is different from the protostellar evolution which is not sensitive to the accretion history \citep[see, e.g.,][]{Hosokawa+11}.

The protostellar evolution is governed by accretion and nuclear burning (see Appendices\,B and C of \citetalias{Kunitomo+17}).
Therefore with the same $\xi$ and $X\sub{D}$ values, stellar structures at a given mass are almost the same.
By contrast, in the pre-MS phase, since accretion rate is small (or zero), radiative cooling dominates.

Therefore, we conclude that our discussions on the comparisons with observations in Sects.\,\ref{sec:obs}--\ref{sec:lamboo} are not affected by the uncertainties in metallicity, but the variation in $t\sub{disk}$ has a significant impact on $\Zs$. We may consider that a relatively long-lived disk such as the one chosen in our fiducial case will maximize the effect of planet formation on stellar metallicity. Larger values are possible however, for disks which are extremely long lived as the 20\,Myr case depicted here.

\section{Evolutionary models of stellar interiors} \label{app:table}

The evolutionary models of stellar interiors shown in this paper may be obtained freely from the CDS (see the footnote in page 1).
The input parameters are $\xi=[0,0.5]$, $X\sub{D}=20$ or 28\,ppm, the uniform heat distribution or $m\sub{ke}=0.1$ and $\Mfin=[0.5\,\Msun,1.5\,\Msun]$.
We adopt a solar metallicity $\Zini=0.01966291$ to match the present Sun's radius and luminosity and $\Za=\Zini$.
The models differ slightly from those of \citetalias{Kunitomo+17}\footnote{http://cdsarc.u-strasbg.fr/viz-bin/qcat?J/A+A/599/A49}, due to different hypotheses for the accretion rate and element diffusion.

The table includes $\MCZeff$ (only between 0.3\,Myr and 10\,Myr, see Sect.\,\ref{sec:MCZeff}), in addition to the temporal evolutions of the mass of an outer convective zone, $\MCZ$.
From this and Eq.\,\ref{eq:simpleZs}, the surface metallicity, $\Zs$, of stars being affected by planet formation can be estimated as
\begin{equation} \label{eq:Zstp}
\Zs({t\sub{p}})=
\frac
{\MCZeff\,{\Zini} + \dM\,\Za}
{\MCZeff + \dM}
\ ,
\end{equation}
where the planet formation time, $t\sub{p}$, indicates the time when $\Za$ changes from an initial value $\Zini$ to a lower value and $\dM(t\sub{p})=\Mfin-\Mstar(t\sub{p})$ is the mass to be accreted.
Readers can evaluate consequences of planet formation on stellar surface composition using the results in Table\,\ref{tab:online} and Eq.\,\eqref{eq:Zstp}.

   \begin{table*}[!tb]
      \caption[]{Stellar evolutionary models.}
         \label{tab:online} \small
          \centering
         \begin{tabular}{cccc|ccccccc} 
            \hline
            \hline
            \noalign{\smallskip}
            \multicolumn{4}{c|}{Settings} & \multicolumn{7}{c}{Results}\\
            \noalign{\smallskip}
            $\xi$ & $X\sub{D}$ & $m\sub{ke}\tablefootmark{a}$ & $\Mfin$   & Age  & $\Mstar$  & $\MCZ$    & $\MCZeff\tablefootmark{b}$ & $R_{\star}$ & $L_{\star}$ & $\teff$  \\
                  & [ppm]      &                              & $[\Msun]$ & [yr] & $[\Msun]$ & $[\Msun]$ & $[\Msun]$ & $[\Rsun]$   & $[\Lsun]$   & [K]  \\
            \noalign{\smallskip}
            \hline
            \noalign{\smallskip}
\tt 0.1 & \tt 28  & \tt  -1.0  & \tt  1.0  & \tt   0.0E+00   & \tt   1.00000E-02  & \tt    1.00000E-02   & \tt  -1.00000E+00  & \tt   1.51399E+00     & \tt 5.44190E-02   & \tt   2.21056E+03\\
\tt 0.1  & \tt  28  & \tt  -1.0  & \tt  1.0   & \tt  1.0E+03   & \tt   1.99973E-02  & \tt    4.35589E-03   & \tt  -1.00000E+00    & \tt  1.13449E+00    & \tt  3.29172E-02   & \tt   2.30986E+03\\
            \noalign{\smallskip}
              \multicolumn{11}{c}{$\vdots$}\\
            \noalign{\smallskip}
              \multicolumn{11}{c}{\textit{Continued}}\\
            \noalign{\smallskip}
            \hline
            \noalign{\smallskip}
         \end{tabular}
         \tablefoot{
         Accretion history described in Sect.\,\ref{sec:method-Mdot} with $\Za=\Zini$ is assumed.
         A complete table is available at CDS.
         $^{(a)}$ 
         Although $m\sub{ke}$ is defined from 0 to 1, here for simplicity $m\sub{ke}=-1$ indicates a uniform heat distribution (Sect.\,\ref{sec:method-SE}). 
         $^{(b)}$ 
         $\MCZeff$ is available only between 0.3 and 10\,Myr and is set to $-1$ if not available.
                  }
        \normalsize
            \end{table*}

   \begin{table}[!tb]
      \caption[]{List of evolutionary models in the online table.}
         \label{tab:online2} \small
          \centering
         \begin{tabular}{ccccc} 
            \hline
            \hline
            \noalign{\smallskip}
            Model & $\xi$ & $X\sub{D}$ & $m\sub{ke}$ & $\Mfin$   \\
             &                    & [ppm]         &                     & $[\Msun]$  \\
            \noalign{\smallskip}
            \hline
            \noalign{\smallskip}
             1 & 0.5 & 28 & $-1$ & 0.5--1.5      \\
             2 & 0.1 & 28 & $-1$ & 0.5--1.5      \\
             3 & 0    & 28 & $-1$ & 0.5--1.5      \\
             4 & 0.1 & 28 & 0.1  & 0.5--1.5      \\
             5 & 0.5 & 20 & $-1$ & 0.5--1.5      \\
             6 & 0.1 & 20 & $-1$ & 0.5--1.5      \\
             7 & 0    & 20 & $-1$ & 0.5--1.5      \\
            \noalign{\smallskip}
            \hline
            \noalign{\smallskip}
         \end{tabular}
         \tablefoot{Model 2 corresponds to our fiducial case. }
        \normalsize
            \end{table}

\end{appendix}
\end{document}